\documentclass[aps,prb,twocolumn,showpacs,superscriptaddress]{revtex4}
\usepackage{amssymb}
\usepackage{graphicx}

\begin{document}
\title{Multi-gap superconductivity and Shubnikov-de Haas oscillations in single crystals of the layered boride OsB$_2$}
\author{Yogesh Singh}
\altaffiliation[Present address: ]{I. Physikalisches Institut, Georg-August-Universit\"at G\"ottingen, D-37077, G\"ottingen, Germany}
\author{C. Martin}
\altaffiliation[Present address: ]{Department of Physics, B2126 New Physics Building, University of Florida, Gainesville, Florida 32611, USA}
\author{S. L. Bud'ko}
\affiliation{Ames Laboratory and Department of Physics and Astronomy, Iowa State University, Ames, Iowa 50011, USA}
\author{A. Ellern}
\affiliation{Department of Chemistry, Iowa State University, Ames, Iowa 50011, USA}
\author{R. Prozorov}
\author{D. C. Johnston}
\affiliation{Ames Laboratory and Department of Physics and Astronomy, Iowa State University, Ames, Iowa 50011, USA}
\date{\today}

\begin{abstract} Single crystals of superconducting OsB$_2$ [$T_{\rm c} = 2.10(5)$~K] have been grown using a Cu-B eutectic flux.  We confirm that OsB$_2$ crystallizes in the reported orthorhombic structure (space group \emph{Pmmn}) at room temperature.  Both the normal and superconducting state properties of the crystals are studied using various techniques.  Heat capacity versus temperature $C(T)$ measurements yield the normal state electronic specific heat coefficient $\gamma = 1.95(1)$~mJ/mol~K$^2$ and the Debye temperature $\Theta_{\rm D} = 539(2)$~K\@.  The measured frequencies of Shubnikov-de Haas oscillations are in good agreement with those predicted by band structure calculations.  Magnetic susceptibility $\chi(T,H)$, electrical resistivity $\rho(T)$ and $C(T,H)$ measurements ($H$ is the magnetic field) demonstrate that OsB$_2$ is a bulk low-$\kappa$ [$\kappa(T_{\rm c}) = 2(1)$] Type-II superconductor that is intermediate between the clean and dirty limits [$\xi(T=0)/\ell = 0.97$)] with a small upper critical magnetic field $H_{\rm c2}(T = 0) = 186(4)$~Oe.  The penetration depth is $\lambda(T = 0) =  0.300~\mu$m.  An anomalous (not single-gap BCS) $T$ dependence of $\lambda$ was fitted by a two-gap model with $\Delta_1(T = 0)/k_{\rm B}T_{\rm c} = 1.9$ and $\Delta_2(T = 0)/k_{\rm B}T_{\rm c} = 1.25$, respectively.  The discontinuity in the heat capacity at $T_{\rm c}$, $\Delta C/\gamma T_{\rm c} = 1.32$, is smaller than the weak-coupling BCS value of 1.43, consistent with the two-gap nature of the superconductivity in OsB$_2$.  An anomalous increase in $\Delta C$ at $T_{\rm c}$ of unknown origin is found in finite $H$; e.g., $\Delta C/\gamma T_{\rm c} \approx 2.5$ for $H \approx 25$~Oe.  
\end{abstract}
\maketitle

\section{Introduction}
\label{sec:INTRO}
Although multigap superconductivity was first addressed theoretically by Suhl \emph{et al.} in 1959,\cite{suhl1959} and the first experimental observation of the possible existence of two distinct superconducting gaps was made in 1980 using tunneling measurements on Nb-doped SrTiO$_3$,\cite{Binnig1980} the subject of multigap superconductivity has only recently gained impetus after it was established that several unusual superconducting properties of MgB$_2$ could be explained within a two-gap superconductivity scenario.\cite{Bouquet2001,choi2002}  There are now several other candidates for multi-gap superconductivity like NbSe$_2$ (Ref.~\onlinecite{Boaknin2003}), $R$Ni$_2$B$_2$C ($R =$~Lu, Y) (Ref.~\onlinecite{Shulga1998}), Lu$_2$Fe$_3$Si$_5$ (Refs.~\onlinecite{Nakajima2008}, \onlinecite{Gordon2008}), and Sr$_2$RuO$_4$.\cite{Maeno2001} 

In multigap superconductors distinct superconducting gaps exist on different disconnected parts (sheets) of the Fermi surface (FS) although the interband pairing leads to a single critical temperature $T_{\rm c}$.\cite{suhl1959,Kresin1990}  Most superconductors show multi-band conduction, but due to interband pairing the gap has the same magnitude on all bands.  When interband pairing is weak then the gaps on different sheets of the FS can have significantly different magnitudes.  This can lead to anomalous behavior in the temperature-dependent heat capacity, upper critical magnetic field $H_{\rm c2}$, and penetration depth $\lambda$ measurements.\cite{Bouquet2001,Shulga1998,Fisher2003,Manzano2002}  

\begin{figure}[t]
\includegraphics[width=3.in]{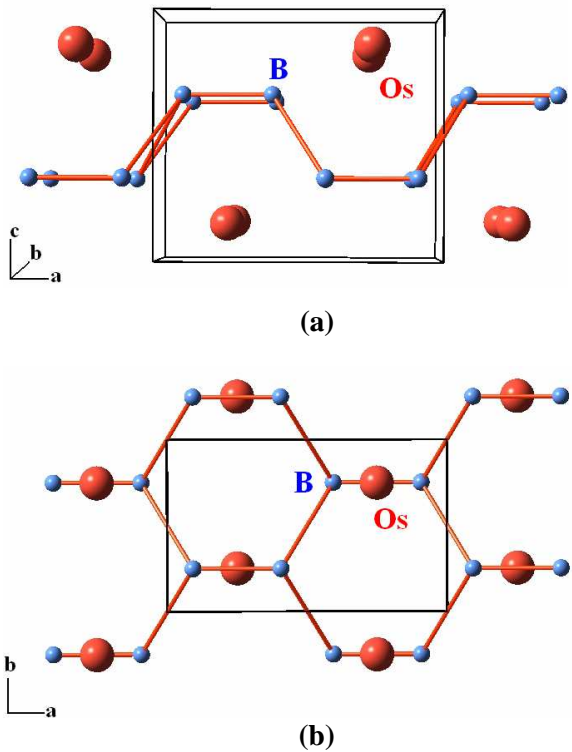}
\caption{(Color online) The crystal structure of OsB$_2$ viewed at a slight angle from the $b$~axis.  The Os atoms are shown as large (red) spheres while the B atoms are shown as the small (blue) spheres.  A single unit cell (shown) contains two formula units.  (b) Projection of the OsB$_2$ structure onto the $ab$~plane. 
\label{Fig-structure}}
\end{figure}

The compound OsB$_2$ has a layered crystal structure qualitatively similar to that of MgB$_2$, except that the B layers are corrugated in OsB$_2$ instead of flat as in MgB$_2$.\cite{Nagamatsu2001}  The crystal strucure of OsB$_2$ is shown in Fig.~\ref{Fig-structure}.  Figure~\ref{Fig-structure}(a) shows the crystal structure of OsB$_2$ viewed at a slight angle from the $b$~axis.  Figure~\ref{Fig-structure}(b) shows the structure projected on the $ab$~plane.  Along the $c$~axis the boron layers lie between two planar transition metal layers which are offset along the \emph{ab}-plane.  We have recently reported\cite{Singh2007} several anomalous behaviors for polycrystalline samples of the layered superconductor OsB$_2$ which has a superconducting transition temperature $T_{\rm c} = 2.1$~K.\cite{Vandenberg1975}  These unusual behaviors include a reduced specific heat discontinuity at $T_{\rm c}$ in some samples and a magnetic field penetration depth versus temperature $T$ dependence that was consistent with two-gap superconductivity.  We also observed a positive curvature in the $T$ dependence of the upper critical magnetic field $H_{\rm c2}$. To gain further insights into these interesting behaviors, measurements on single crystals are needed.    

Herein we report the growth of OsB$_2$ single crystals, and structure, isothermal magnetization, dynamic and static magnetic susceptibility, specific heat, electrical resistivity, magnetic field penetration depth, and Shubnikov-de Haas (SdH) oscillation measurements on the crystals to characterize their superconducting and normal state properties.  Following a description of the experimental details in Sec.~\ref{sec:EXPT}, the experimental results are given in Sec.~\ref{results}.  A summary of the results and our conclusions are given in Sec.~\ref{sec:CON}, including a list in Table~\ref{PropSumm} summarizing the parameters characterizing the physical properties that we obtained.

\begin{figure}[t]
\includegraphics[width=2.1in,angle=90]{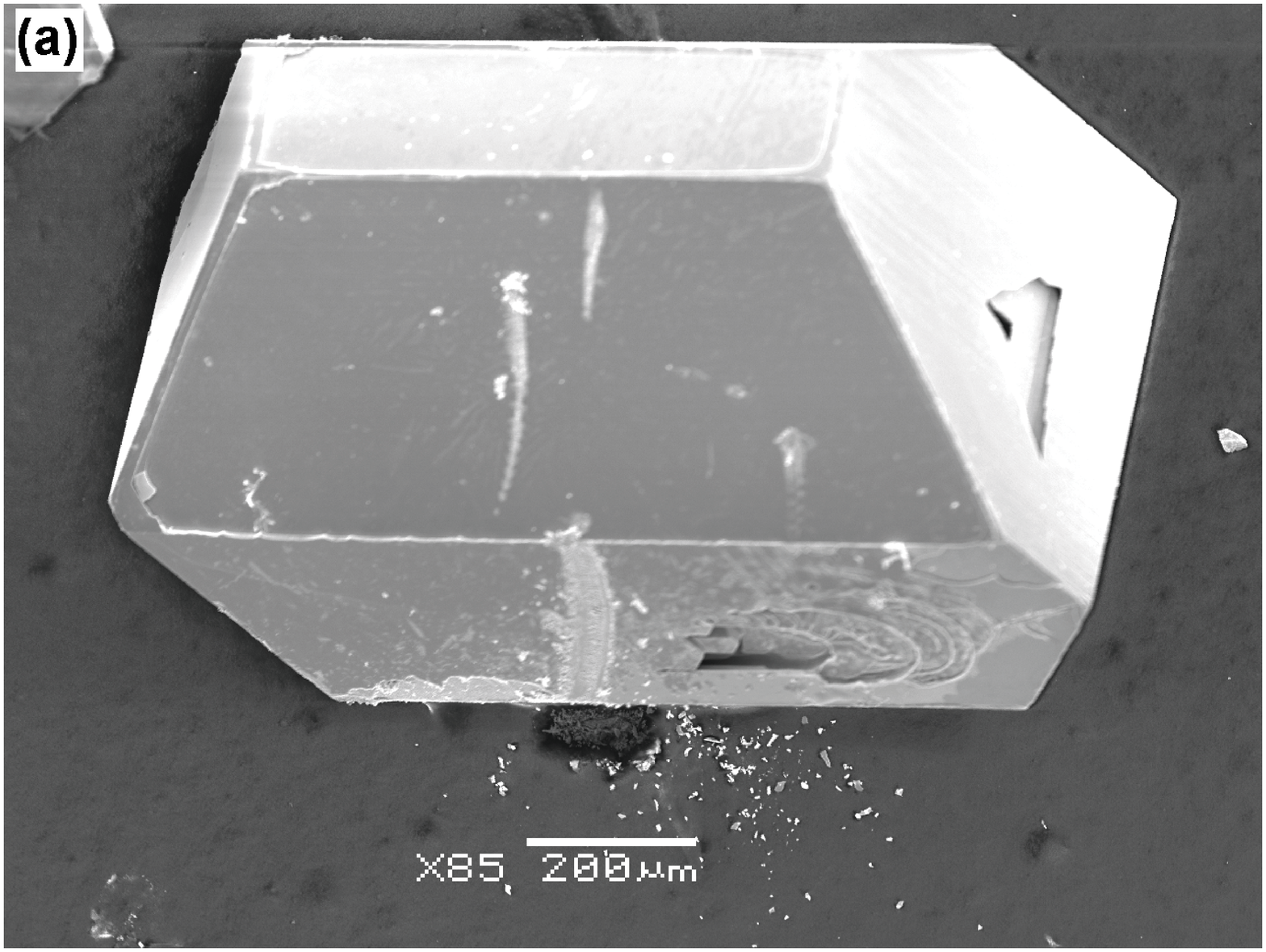}
\vskip 0.5cm
\includegraphics[width=3.1in]{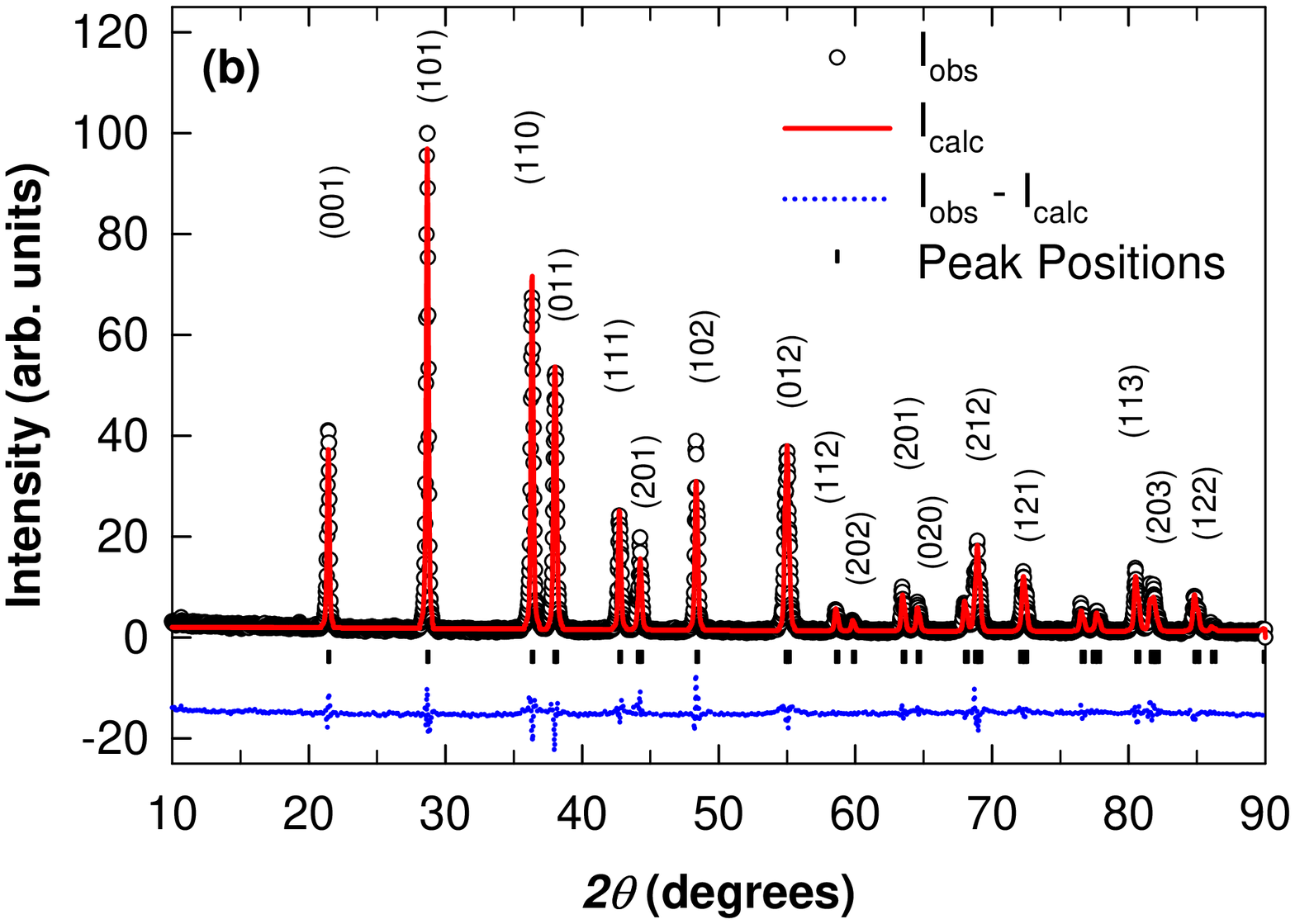}
\caption{(Color online) (a) Scanning electron microscope image of a typical OsB$_2$ crystal.  (b) Rietweld refinement of the powder X-ray diffraction data of crushed OsB$_2$ crystals.  The open black circles represent the observed X-ray pattern, the solid red line represents the fitted pattern, the dotted blue line represents the difference between the observed and calculated intensities and the vertical black bars represent the peak positions.  
\label{Figxrd}}
\end{figure}

\section{EXPERIMENTAL DETAILS}
\label{sec:EXPT}
Single crystals of OsB$_2$ were grown with a high temperature solution growth method using Cu-B as the flux.  First, a polycrystalline sample of OsB$_2$ was prepared by arc-melting Os powder (99.95\%, Alfa Aesar) and B chunks (99.5\%, Alfa Aesar) taken in stoichiometric ratio.  A Cu-B binary alloy was then prepared at the eutectic composition Cu$_{0.87}$B$_{0.13}$ by arc-melting.  For crystal growth the arc-melted OsB$_2$ sample ($\sim$ 0.5~g) was placed in a 2~mL Al$_2$O$_3$ crucible.  About 5~g Cu-B flux was placed on top of the OsB$_2$ ingot.  The crucible with a lid was placed in a vertical tube furnace which was then evacuated and purged with high purity Ar gas repeatedly ($\approx$~10 times) after which the growth was started in a flow ($\approx 60$~cc/min) of Ar.  The furnace was heated to 800~$^{\circ}$C in 30~min, then heated to 1450~$^{\circ}$C in 6~hrs and held at this temperature for 6~hrs.  The furnace was then cooled to 1020~$^{\circ}$C at a rate of 2~$^{\circ}$C/hr and then rapidly cooled to room temperature.  Well-formed crystals with flat facets were obtained after the Cu-B flux had been dissolved in dilute nitric (HNO$_3$) acid.  A scanning electron micrograph of a typical crystal is shown in Fig.~\ref{Figxrd}(a).

Some single crystals were crushed for powder X-ray diffraction (XRD) measurements.  The XRD patterns were obtained using a Rigaku Geigerflex diffractometer with Cu K$\alpha$ radiation, in the 2$\theta$ range from 10 to 90$^\circ$ with a 0.02$^\circ$ step size. Intensity data were accumulated for 5~s per step.   

For single crystal structure determination, a well-shaped crystal ($0.27 \times 0.18 \times 0.15$ mm$^3$) was selected.  The data collection for the crystal was performed using a Bruker Apex II instrument with Cu~K$\alpha$ radiation at $T = 100$~K and was solved with latest version of the Apex software package which is reliable for a combination of numerical and multi-scan absorption correction.  The initial cell constants were obtained from three series of $\omega$ scans at different starting angles.  Each series consisted of 30 frames collected at intervals of 0.3$^{\circ}$ in a 10$^{\circ}$ range about $\omega$ with the exposure time of 5~s per frame.  The obtained reflections were successfully indexed by an automated indexing routine built in the Apex program.  The final cell constants were calculated from a set of strong reflections from the actual data collection.  The data were collected using the full sphere routine by collecting 20 sets of frames with 1 degree scans in $\omega$ with an exposure time of 5~s per frame.  This data set was corrected for Lorentz and polarization effects.  The absorption correction was a combination of a numerical one based on a face indexing and an additional correction based on fitting a function to the empirical transmission surface as sampled by multiple equivalent measurements\cite{Blessing1995} using the Apex software.\cite{SHELXTL}

The temperature dependences of the dc magnetic susceptibility and isothermal magnetization were measured using a commercial Superconducting Quantum Interference Device (SQUID) magnetometer (MPMS5, Quantum Design).  The resistivity and heat capacity were measured using a commercial Physical Property Measurement System (PPMS, Quantum Design).  The resistivity was measured using a four-probe technique with a current of 5~mA along the $b$ axis.  The dynamic susceptibility was measured between 0.5~K and 2.6~K using a 10~MHz tunnel-diode driven oscillator (TDO) circuit with a volume susceptibility sensitivity $\Delta\chi\approx 10^{-8}$.\cite{prozorov2006r}  The details of the measurement and the extraction of magnetic susceptibility and penetration depth from TDO measurements have been described in our previous work.\cite{Singh2007}

\section{\label{results} Results}
\subsection{Crystal Structure of OsB$_2$}
\label{sec:RES-structure}

\begin{table}
\caption{\label{CrystalData} Crystal data and structure refinement of OsB$_2$.  Here $R1 = \sum$$\mid$$\mid$$F$$_{\rm obs}$$\mid$~$-$~$\mid$$F$$_{\rm calc}$$\mid$$\mid$/$\sum$$\mid$$F$$_{\rm obs}$$\mid$ and $wR2 = (\sum$[ $w$($\mid$$F$$_{\rm obs}$$\mid$$^2$ $-$ $\mid$$F$$_{\rm calc}$$\mid$$^2$)$^2$]/$\sum$[ $w$($\mid$$F$$_{\rm obs}$$\mid$$^2$)$^2$])$^{1/2}$, where $F$$_{\rm obs}$ is the observed structure factor and $F$$_{\rm calc}$ is the calculated structure factor.}

\begin{ruledtabular}
\begin{tabular}{ll}
Temperature & 100(2) K\\
Crystal system, space group & Orthorhombic, \emph{Pmmn}\\
Unit cell parameters & $a$ = 4.6729(3) \AA,\\
& $b$ = 2.8702(2) \AA,\\
& $c$ = 4.0792(3) \AA\\
Unit cell volume & 54.711(7) \AA$^3$ \\
$Z$ (formula units per unit cell)& 2 \\
Molar volume & 16.474(2) cm$^3$/mol \\
Density (calculated) & 12.858 Mg/m$^3$ \\
Absorption coefficient & 212.327 mm$^{-1}$ \\
F(000) & 172 \\
Data / restraints / parameters & 65 / 12 / 12 \\
Goodness-of-fit on $F$$^2$ & 1.015 \\
Final $R$ indices [$I > 2\sigma(I)$] & $R1 = 0.0294$ \\
& $wR2 = 0.0736$ \\ 
Extinction coefficient & 0.016(3) \\
\end{tabular}
\end{ruledtabular}
\end{table}

\begin{table}
\caption{\label{atomicpositions}
Atomic coordinates and anisotropic displacement parameters for OsB$_2$ in space group \emph{Pmmn} (second setting).  The Os atoms occupy Wyckoff 2\emph{a} $\left(\frac{1}{4},\frac{1}{4},z\right)$ positions and the B atoms occupy 4\emph{f} $\left(x,\frac{1}{4},z\right)$ positions.  $U_{11}$, $U_{22}$ and $U_{33}$ are the anisotropic thermal parameters in units of \AA$^2$ defined within the thermal parameter of the intensity as $e^{-2\pi^2(h^2a^2U_{11} + k^2b^2U_{22} + l^2c^2U_{33})}$.}

\begin{ruledtabular}
\begin{tabular}{c|cccccc}
atom & \emph{x} & \emph{y} & \emph{z} & \emph{U}$_{11}$ & \emph{U}$_{22}$ & \emph{U}$_{33}$ \\ \hline  
Os & 1/4 & 1/4 & 0.1527(2) & 0.003(1) & 0.004(1) &	0.003(1)\\  
B &0.049(6) & 1/4 & 0.359(4) & 0.004(8) & 0.001(8) & 0.006(8)\\ 
\end{tabular}
\end{ruledtabular}
\end{table}

The powder XRD pattern for crushed single crystals of OsB$_2$ is shown in Fig.~\ref{Figxrd}(b).  All the lines in the X-ray pattern could be indexed to the known orthorhombic \emph{Pmmn} (No.~59) structure and a Rietveld refinement\cite{Rietveld} of the X-ray pattern, shown in Fig.~\ref{Figxrd}(b), gave the lattice parameters \emph{a}~=~4.6855(6)~\AA, \emph{b}~=~2.8730(3)~\AA~and \emph{c}~=~4.0778(4)~\AA.  These values are in very good agreement with our previously reported values [$a$~=~4.6851(6)~\AA, \emph{b}~=~2.8734(4)~\AA, and \emph{c}~=~4.0771(5)~\AA ] for a polycrystalline sample.\cite{Singh2007}  

Single crystal XRD data were obtained at $T = 100$~K\@.  The systematic extinctions of peaks in the XRD data were consistent with the space group \emph{Pmmn},\cite{SHELXTL} in agreement with earlier reports from single crystal and powder XRD measurements on OsB$_2$.\cite{Roof1962}  The positions of the atoms were found by  direct methods and were refined in full-matrix anisotropic approximation.  Some parameters obtained from the single crystal structure refinement are given in Table~\ref{CrystalData} and the final atomic positions and anisotropic thermal parameters are given in Table~\ref{atomicpositions}.

\subsection{Electrical Resistivity}
\label{sec:RES}
\begin{figure}[t]
\includegraphics[width=3.in]{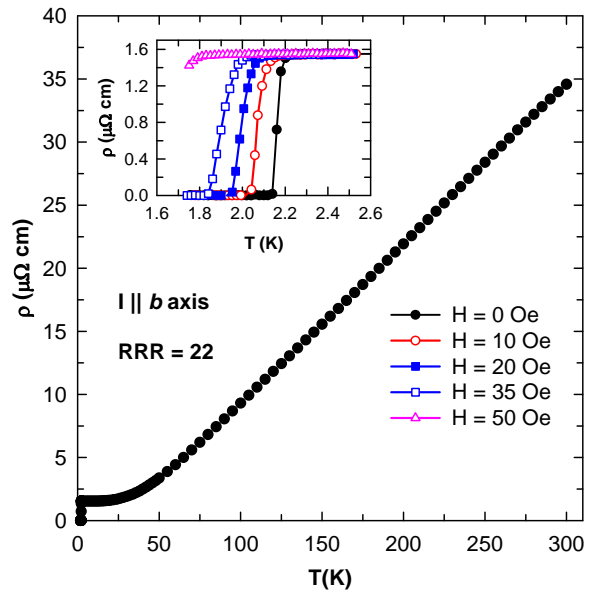}
\caption{(Color online) Electrical resistivity $\rho$ for a single crystal of OsB$_2$ versus temperature \emph{T} with current $I$~=~5~mA along the $b$ axis and with applied magnetic field $H$~=~0~Oe.  The inset shows the $\rho(T)$ data between $T = 1.7$ and~2.6~K measured in various applied magnetic fields $H$ as indicated.
\label{Fig-OsB2RES}}
\end{figure}

The electrical resistivity $\rho$ versus $T$ of a single crystal of OsB$_2$ from 1.75~K to 300~K measured in zero applied magnetic field $H$ and with a current $I$~=~5~mA applied along the $b$ axis, is shown in Fig.~\ref{Fig-OsB2RES}.  The $\rho(T)$ shows metallic behavior with an approximately linear decrease in resistivity on cooling from room temperature to 50~K\@.  This behavior is similar to that observed earlier for a polycrystalline sample.\cite{Singh2007}  At low temperatures $\rho$ becomes only weakly temperature dependent and reaches a residual resistivity $\rho_0 = 1.55~\mu\Omega$~cm just above 2.2~K\@ as seen in the inset of Fig.~\ref{Fig-OsB2RES}.  The large residual resistivity ratio RRR~=~$\rho$(300~K)/$\rho_0$~=~22 indicates a well crystallized sample.

The inset of Fig.~\ref{Fig-OsB2RES} shows the low $T$ data measured in various $H$.  The $\rho$ in $H = 0$ drops abruptly below 2.20~K and reaches zero by 2.14~K, as highlighted in the inset of Fig.~\ref{Fig-OsB2RES}.  This superconducting transition was observed earlier by us for a polycrystalline sample,\cite{Singh2007} consistent with the original report in 1975 of superconductivity in OsB$_2$ by Vandenberg et al.\cite{Vandenberg1975}  As expected, the superconducting transition shifts to lower $T$ with increasing $H$.  These data were used to determine the upper critical magnetic field $H_{\rm c2}(T)$ which will be discussed later.  In particular, for each applied field $H$, this $H$ is taken to be $H_{\rm c2}$ for the temperature at which the resistance drops to zero.

\begin{figure}[t]
\includegraphics[width=2.5in, angle=-90]{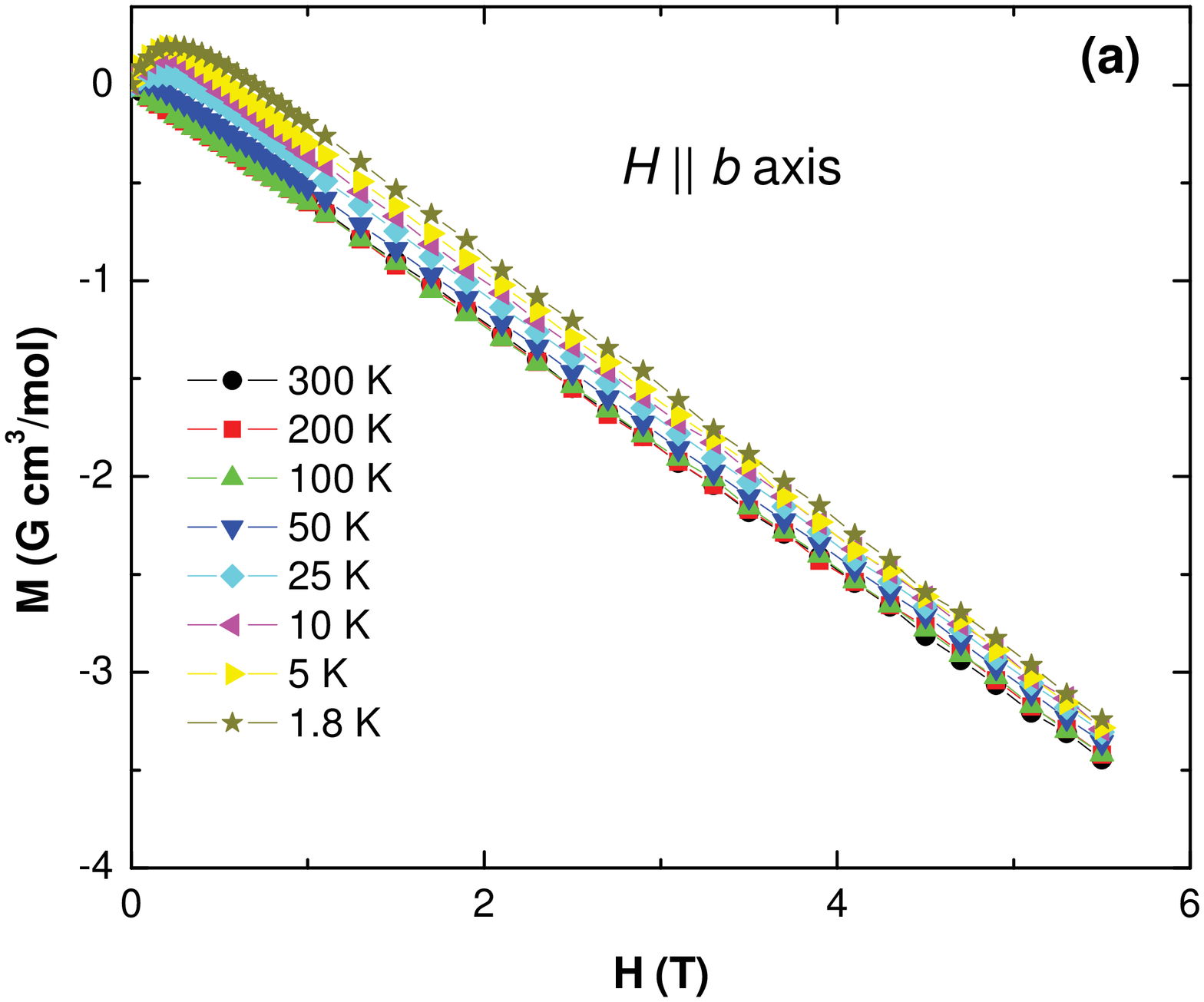}\vspace{0.2in}
\includegraphics[width=2.5in, angle=-90]{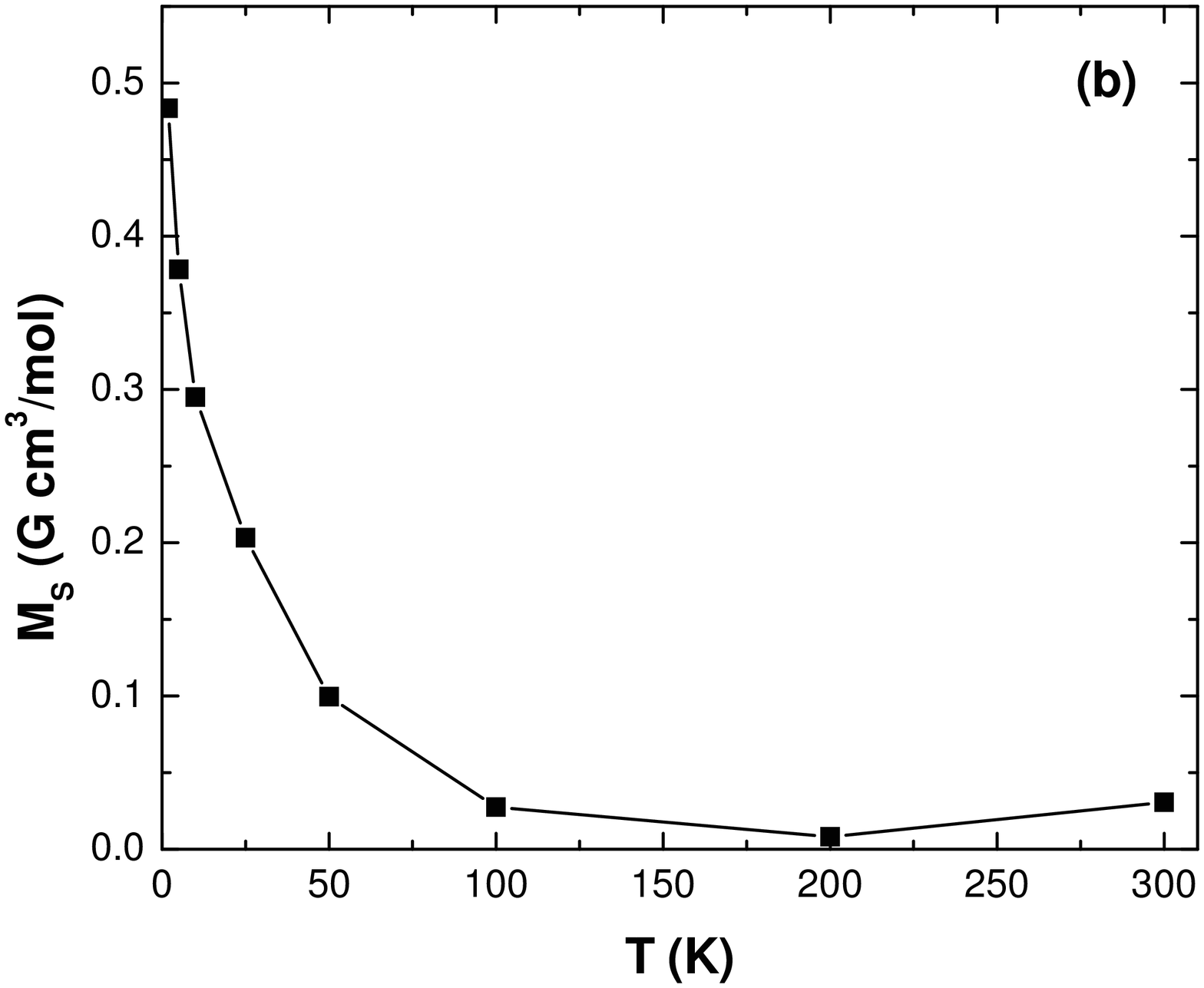}
\caption{(Color online) (a) Isothermal magnetization $M$ versus magnetic field $H$ at various temperatures $T$ with $H$ applied along the $b$ axis.  (b) Saturation magnetization $M_{\rm S}$ versus $T$ obtained by fitting the $M(H)$ data above $H = 2$~T at various $T$ by the expression $M(H,T) = M_S(T) + \chi(T) H$.    
\label{Fig-M(H)}}
\end{figure}

\subsection{Isothermal Magnetization and Magnetic Susceptibility}
\label{sec:RES-normalstate-susceptibility}
\subsubsection{Normal State}

The isothermal magnetization $M$ versus $H$ applied along the $b$ axis is shown in Fig.~\ref{Fig-M(H)}(a) at various $T$.  At high $T > 100$~K, $M$ is diamagnetic and proportional to $H$ with a slope that is almost constant between $T = 100$ and $T = 300$~K\@.  At lower $T$, $M$ initially increases with $H$ towards a positive value before showing saturation at a field of about 500~Oe.  For higher $H$, $M$ turns over and becomes diamagnetic.  For $H > 1$~T, $M$ is linear in $H$ with a nonzero $y$-intercept.  A similar $M(H)$ behavior is also observed in measurements with $H$ applied along the $a$ and $c$ axes and is consistent with the presence of a small amount of paramagnetic and/or ferromagnetic impurities in the sample.  Contributions from paramagnetic impurities are also observed at low temperatures in our normal state magnetic susceptibility $\chi$ measurements via a Curie-Weiss-like upturn in Fig.~\ref{Figchi_normal} below.  The $M(H)$ data indicate that the impurity contribution saturates at high $H$.  We extract the saturation magnetization $M_{\rm S}(T)$ by fitting the $M(H,T)$ by the expression $M(H,T) = M_{\rm S}(T) + \chi_{\rm int} H$, where $\chi_{\rm int}$ is the intrinsic susceptibility.  The $M_{\rm S}(T)$ data so obtained are shown in Fig.~\ref{Fig-M(H)}(b).

The normal state $\chi\equiv M/H$ versus $T$ for OsB$_2$, measured between 1.8~K and 300~K with $H = 3$~T applied along the $a$, $b$, and $c$ axes, is shown in Fig.~\ref{Figchi_normal}.  The powder average susceptibility $\chi(T)$ and the $\chi(T)$ for a polycrystalline sample from Ref.~\onlinecite{Singh2007} are also shown in Fig.~\ref{Figchi_normal}.  The $\chi(T)$ for single crystalline OsB$_2$ is weakly temperature dependent between 50~K and 300~K\@.  The upturn at low temperatures is most likely due to the presence of small amounts of paramagnetic impurities as mentioned above.  The solid curve through the powder average $\chi(T)$ data is a fit by the expression $\chi(T) = \chi_{\rm int} + {C\over T-\theta}$.  The fit gave the values $\chi_{\rm int} = -4.56(3) \times 10^{-5}$~cm$^3$/mol, $C = 1.43(1) \times 10^{-4}$~cm$^3$~K/mol, and $\theta = -12.1(5)$~K\@.  This value of $C$ is equivalent to about 0.04~mol\% of spin-1/2 impurities with a $g$-factor $g = 2$.  The large negative value of $\theta$ is probably due, at least in part, to saturation of the paramagnetic impurities by the relatively high 3~T field.  The powder average value at 300~K is $\bar{\chi}$(300 K) = ${\rm -4.50 \times 10^{-5}\ cm^3/mol}$.  A diamagnetic susceptibility for a transition metal compound is rare, but not unprecedented.\cite{Singh2007}

Fig.~\ref{Figchi_normal} also shows the intrinsic susceptibility $\chi_{\rm int}(T)$ obtained by correcting the powder averaged susceptibility for the presence of ferromagnetic and/or paramagnetic impurities as discussed above.  Here we have assumed that $M_{\rm S}(T)$ is isotropic so that we can use the $M_{\rm S}(T)$ data obtained from $M(H,T)$ data for the $b$ axis to correct the powder averaged susceptibility.

\begin{figure}[t]
\includegraphics[width=2.5in, angle=-90]{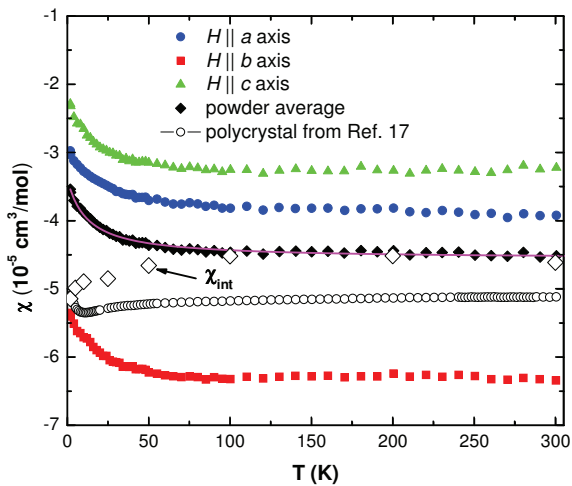}
\caption{ (Color online) Magnetic susceptibility $\chi$ versus temperature \emph{T} for a single crystal of OsB$_2$ with a magnetic field $H = 3$~T applied along the $a$, $b$, and $c$ axes.  The powder average $\chi(T)$ and the $\chi(T)$ for a polycrystalline sample (from Ref.~\onlinecite{Singh2007}) are also shown.  The solid curve through the powder average $\chi(T)$ data is a fit by the expression $\chi(T) = \chi_{\rm int} + {C\over T-\theta}$.  The intrinsic susceptibility obtained by correcting the powder averaged $\chi(T)$ for the ferromagnetic impurity contribution $M_{\rm S}(T)$, $\chi_{\rm int}(T) = \chi(T) -M_{\rm S}(T)/H $ is shown as open diamond symbols.   
\label{Figchi_normal}}
\end{figure}

The $\chi$ along different crystallographic directions is anisotropic with the value (averaged between $T = 100$~K and 300~K) along the $b$ axis ($\approx -6.25 \times 10^{-5}$~cm$^3$/mol) being much smaller than the susceptibility along the $a$ ($\approx -3.75 \times 10^{-5}$~cm$^3$/mol) or $c$ ($\approx -3.25 \times 10^{-5}$~cm$^3$/mol) axes which are quite similar.  The similarity of the $\chi$ values along the $a$ and $c$ axes is surprising given the layered nature of the crystal structure which is built up of alternating Os and B layers in the $a$-$b$ plane that are stacked along the $c$ axis.  However, theoretical Fermi surface (FS) calculations have shown that there are quasi-one-dimensional tubular structures running along the $b$ axis, and the FSs along the $a$ and $c$ axes are quite similar.\cite{Hebbache-FS}

As described in Ref.~\onlinecite{Singh2007}, one can estimate the paramagnetic Pauli spin susceptibility $\chi_{\rm P}$ from the intrinsic susceptibility $\chi_{\rm int}$ according to

\begin{equation}
\chi_{\rm int} = \chi_{\rm orb} + \chi_{\rm P},
\label{EqChiChiorbChip}
\end{equation}
where $\chi_{\rm orb}$ is the total orbital susceptibility, which includes the diamagnetic core contribution, the paramagnetic Van Vleck contribution, and the Landau diamagnetic contribution from the conduction electrons. In Ref.~\onlinecite{Singh2007}, we estimated $\chi_{\rm orb} = -{\rm 7.8 \times 10^{-5}\ cm^3/mol}$.  However, the accuracy of this estimate for $\chi_{\rm orb}$ is unknown (see also below).  Using this value of $\chi_{\rm orb}$, our measured $\bar{\chi}_{\rm int}$(300 K) = ${\rm -4.50 \times 10^{-5}\ cm^3/mol}$ and Eq.~(\ref{EqChiChiorbChip}), for our single crystalline OsB$_2$ we obtain a powder average $\chi_{\rm P} = 3.23\times 10^{-5}$~cm$^3$/mol.  

From $\chi_{\rm P}$ one can estimate the density of states at the Fermi level $N(\epsilon_{\rm F})$ for both spin directions using the relation\cite{Kittel}

\begin{equation}
\chi_{\rm P} = \mu_{\rm B}^2 N(\epsilon_{\rm F}) = \left(3.233\times 10^{-5}~{\rm \frac{cm^3}{mol}}\right)N(\epsilon_{\rm F}), 
\label{EqDOSCHIP}
\end{equation}
where $\mu_{\rm B}$ is the Bohr magneton and the equality on the far right-hand side is for $N(\epsilon_{\rm F})$ in units of states/(eV f.u.) for both spin directions, where ``f.u.''\ means ``formula unit.''  Taking the above average value of $\chi_{\rm P}$ for OsB$_2$, we get $N(\epsilon_{\rm F})$~=~$1.00(3)~$states/(eV~f.u.) for both spin directions.  This value is a factor of two larger than the value from our specific heat measurements below as well as from band structure calculations [$N(\epsilon_{\rm F}) = 0.55$~states/(eV~f.u.)],\cite{Hebbachea2006} indicating that our estimate of the orbital susceptibility above is too negative.  Using Eq.~(\ref{EqDOSCHIP}) and the band structure density of states value gives the revised estimate $\chi_{\rm P} = 1.8\times 10^{-5}$~cm$^3$/mol.  Then using the measured $\bar{\chi}_{\rm int}$(300 K) and Eq.~(\ref{EqChiChiorbChip})  yields a revised powder averaged orbital susceptibility $\chi_{\rm orb} = {\rm -6.3 \times 10^{-5}\ cm^3/mol}$.

\begin{figure}[t]
\includegraphics[width=2.25in, angle=-90]{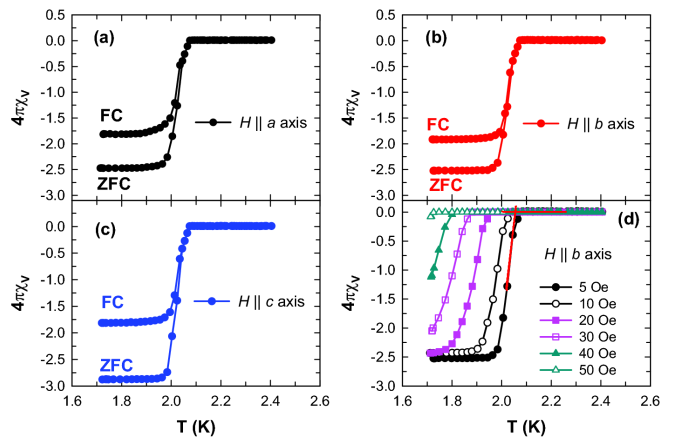}
\caption{(Color online)  Temperature \emph{T} dependence of the zero-field-cooled (ZFC) and field-cooled (FC) dimensionless volume susceptibility $\chi_{\rm v}$ in terms of the superconducting volume fraction (4$\pi\chi_{\rm v}$) of a single crystal of OsB$_2$ with a magnetic field $H = 5$~Oe applied along the (a) $a$ axis, (b) $b$ axis, and (c) $c$ axis.  (d) The $T$ dependence of the ZFC superconducting volume fraction 4$\pi$$\chi_{\rm v}$ of a single crystal of OsB$_2$ measured in various magnetic fields $H$ applied along the $b$ axis.  The construction used to determine $T_{\rm c}(H)$ is illustrated by the red straight line for $H = 5$~Oe.  At low $T$, the 4$\pi\chi_{\rm v}$ values are all more negative than $-1$ due to demagnetization effects.
\label{Figsus}}
\end{figure}

\subsubsection{Superconducting State}
\label{sec:RES-SC-magnetization-acsusceptibility}

The temperature dependence of the anisotropic zero-field-cooled (ZFC) and field-cooled (FC) dimensionless dc volume magnetic susceptibility $\chi_{\rm v}$ of a single crystal of OsB$_2$ measured from 1.7 to 2.5~K is plotted in Figs.~\ref{Figsus}(a), (b), and (c) in a field of 5~Oe parallal to the $a$, $b$ and $c$~axes, respectively, where $\chi_{\rm v} = M_{\rm v}/H$ and $M_{\rm v}$ is the volume magnetization.  Complete diamagnetism in the absence of demagnetization effects corresponds to $\chi_{\rm v} = -1/4\pi$, so the data have been normalized by 1/4$\pi$. A sharp diamagnetic drop in the susceptibility along all three directions, below $T_{\rm c}$~=~2.05~K, signals the transition into the superconducting state.  The large Meissner fraction seen in the FC data for all three directions indicates weak magnetic flux pinning in the crystal.  The data have not been corrected for the demagnetization factors $N_{\alpha}$ ($\alpha = a,\ b,\ c$) which give $-4\pi\chi_{\rm v} = \frac{1}{1-N_{\alpha}}$ for the respective measured value.  From the ZFC data at the lowest temperatures in Fig.~\ref{Figsus}, one obtains $N_a = 0.60$, $N_b = 0.60$ and $N_c = 0.65$, yielding $N_a + N_b + N_c = 1.85$.  This sum is greater than the value of unity expected for an ellipsoid of revolution.  The reason for this discrepancy is not known.

\begin{figure}[t]
\includegraphics[width=2.3in, angle=-90]{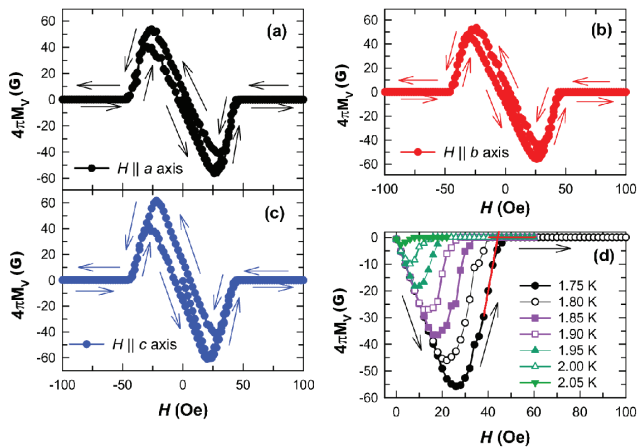}
\caption{(Color online)  Hysteresis loops at 1.7~K of the volume magnetization $M_{\rm v}(H)$ normalized by 1/4$\pi$, versus applied magnetic field \emph{H} applied along the (a) $a$ axis, (b) $b$ axis, and (c) $c$ axis.  The arrows next to the data indicate the direction of field ramping during the measurement.  (d) Normalized magnetization $M_{\rm v}(H)$ at various temperatures $T$ versus $H$ along the $b$ axis.  The construction used to determine $H_{\rm c2}(T)$ is shown by the solid red line for $T = 1.75$~K\@.
\label{FigMH}}
\end{figure}

In Fig.~\ref{Figsus}(d) the temperature dependences of $\chi_{\rm v}$ measured with various $H$ applied along the $b$ axis are shown.  As expected the superconducting transition is suppressed to lower temperatures with increasing $H$.  From these $\chi_{\rm v}(H)$ data the critical field $H_{\rm c2}(T)$ has been estimated using the construction in Fig.~\ref{Figsus}(d), illustrated for $H = 5$~Oe.  The $H_{\rm c2}(T)$ has been determined by fitting a straight line to the data for a given field in the superconducting state just below $T_{\rm c}$ and to the data in the normal state above $T_{\rm c}$ and taking the temperature at which these lines intersect as the $T_{\rm c}$ at that $H$.

The hysteretic volume magnetization $M_{\rm v}$ normalized by 1/4$\pi$ versus $H$ loops measured at $T = 1.7$~K with $H$ applied along the $a$, $b$, and $c$ axes are shown in Figs.~\ref{FigMH}(a), (b), and (c), respectively.  There is a large reversible part in all the $M_{\rm v}$ data recorded with increasing and decreasing $H$ which again indicates very weak magnetic flux pinning in the material.  The $M_{\rm v}(H)$ data recorded at various fixed $T$ with $H$ applied along the $b$ axis are shown in Fig.~\ref{FigMH}(d).  Similar data (not shown) with $H$ along the $a$ and $c$ axes were also recorded.  The initial slope of the $M_{\rm v}(H)$ curves is larger than the value $-1$ expected for perfect diamagnetism, which indicates a nonzero demagnetization factor, consistent with the $4\pi\chi_{\rm v}(T)$ data in Fig.~\ref{Figsus}.  From the $M_{\rm v}(H)$ curves in Fig.~\ref{FigMH}(d) we estimated the critical field $H_{\rm c2}(T)$ from the construction illustrated in Fig.~\ref{FigMH}(d) for $T = 1.75$~K\@.  

\begin{figure}[t]
\includegraphics[width=2.6in, angle=-90]{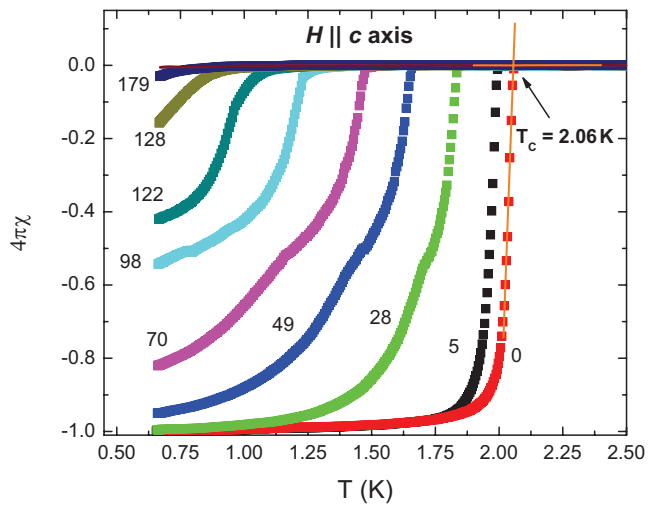}
\caption{(Color online) Dynamic susceptibility $\chi$ normalized by 1/4$\pi$, versus temperature \emph{T} at a frequency of 10~MHz with various applied magnetic fields $H$ in units of Oe.  The $4\pi\chi$ data have been normalized to a minimum value of $-1$ at the lowest $T$.  The construction used to determine $T_{\rm c}(H)$ is shown by the red line for $H = 0$.
\label{FigM(H)-vs-T}}
\end{figure}

The dynamic ac susceptibility $\chi(T)$ measured between 0.6~K and 2.5~K at a frequency of 10~MHz in various $H$ is shown in Fig.~\ref{FigM(H)-vs-T}.  To determine $H_{\rm c2}(T)$ from the data in Fig.~\ref{FigM(H)-vs-T} we fitted a straight line to the data in the normal state and to the data just below $T_{\rm c}$ for a given applied magnetic field and took the value of the \emph{T} at which these lines intersect as the $T_{\rm c}(H)$.  This construction is shown in Fig.~\ref{FigM(H)-vs-T} for the data at \emph{H}~=~0.  By inverting $T_{\rm c}(H)$ we obtain $H_{\rm c2}(T)$.  The $H_{\rm c2}$ has also been obtained in a similar way from the $\chi(T)\equiv M(T)/H$ SQUID magnetometer data (not shown here) between 1.7~K and 2.4~K in various applied magnetic fields.

\subsection{Heat Capacity}
\label{sec:RES-normalstate-heatcap}
\begin{figure}[t]
\includegraphics[width=2.7in,angle=-90]{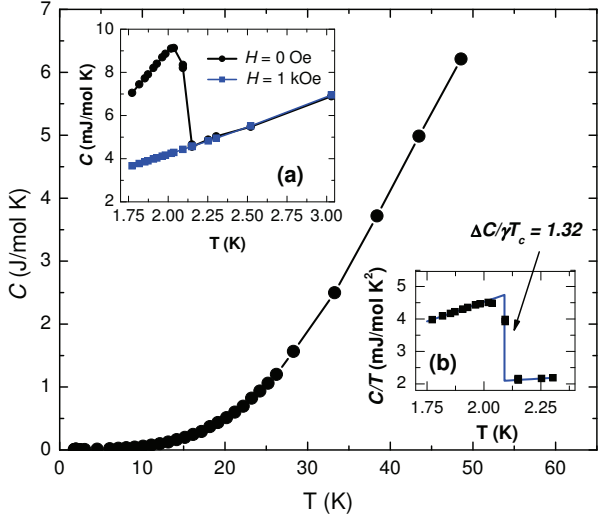}
\caption{Heat capacity \emph{C} versus temperature \emph{T} of a single crystal of OsB$_2$ between 1.75~K and 48~K measured in zero magnetic field $H$.  Inset (a) shows the $C(T)$ data between 1.75~K and 3~K measured with $H = 0$ and $H = 1$~kOe applied along the $c$ axis.  Inset (b) shows the data for $H = 0$ plotted as $C(T)/T$ versus $T$ between $T = 1.75$ and 2.5~K\@.  The solid curve in inset (b) is a construction to estimate the heat capacity jump $\Delta C$ at $T_{\rm c}$.    
\label{Fig(Os,Ru)B2Cp}}
\end{figure}

Figure~\ref{Fig(Os,Ru)B2Cp} shows the heat capacity $C$ versus $T$ data measured on a single crystal of OsB$_2$.  The main panel shows the $C(T)$ data measured in $H = 0$ between $T = 1.75$~K and 48~K\@.  The data below $T = 10$~K could be fitted by the expression $C/T = \gamma + \beta T^2$ where the first term is the contribution from the conduction electrons and the second term is the contribution from the lattice.  The fit (not shown) gave the values $\gamma$~=~1.95(1)~mJ/mol~K$^2$ and $\beta$~=~0.0372(3)~mJ/mol~K$^4$.  From the value of $\beta$ one can estimate the Debye temperature $\theta_{\rm D}$ using the expression \cite{Kittel} 

\begin{equation}
\Theta_{\rm D} = \bigg({12\pi^4Rn \over 5\beta}\bigg)^{1/3} =\left[(1.944\times 10^6)\frac{n}{\beta}\right]^{1/3}, 
\label{EqDebyetemp}
\end{equation}
\noindent
where $R$ is the molar gas constant, $n$ is the number of atoms per formula unit ($n = 3$ for OsB$_2$), and the equality on the far right-hand side is for $\Theta_{\rm D}$ in K and $\beta$ in mJ/mol~K$^4$.  We obtain $\Theta_{\rm D} = 539(2)$~K for OsB$_2$.  The values of $\gamma$, $\beta$, and $\theta_{\rm D}$ obtained above are in very good agreement with the values we reported previously for an unannealed polycrystalline sample.\cite{Singh2007}

The $C(T)$ data below $T$~=~3~K measured in zero and 1~kOe applied field are shown in inset~(a) of Fig.~\ref{Fig(Os,Ru)B2Cp}.  A sharp step-like anomaly at $T$~=~2.1~K is observed in the $H$~=~0~Oe data and confirms the bulk nature of the superconductivity in single crystal OsB$_2$.  The anomaly is suppressed to below $T$~=~1.75~K in a field of $H = 1$~kOe.  The inset (b) shows the $H = 0$~Oe heat capacity plotted as $C(T)/T$ versus $T$ between $T = 1.75$ and 2.5~K\@.  The jump in the specific heat $\Delta C$ at the superconducting transition $T_{\rm c}$ is usually normalized as $\Delta C/\gamma T_{\rm c}$.  From the construction shown as the solid curve through the data in Fig.~\ref{Fig(Os,Ru)B2Cp} inset~(b) we obtain $\Delta C/\gamma T_{\rm c}$~=~1.32.  This value is smaller than the weak-coupling BCS value 1.43.\cite{Tinkham}  Considering the sharp anomaly observed at $T_{\rm c}$ it is unlikely that the smaller value of $\Delta C/\gamma T_{\rm c}$ arises from a distribution of $T_{\rm c}$'s due to inhomogeneties in the sample.  We suggest that the small value of $\Delta C/\gamma T_{\rm c}$ arises from the multi-gap nature of the superconductivity as evidenced from our penetration depth measurements discussed later.

The electron-phonon coupling constant $\lambda_{\rm ep}$ can be estimated in the single-gap superconductivity approximation using McMillan's formula \cite{McMillan1967} which relates the superconducting transition temperature $T_{\rm c}$ to $\lambda_{\rm ep}$, the Debye temperature $\Theta_{\rm D}$, and the Coulumb repulsion constant $\mu^*$, 

\begin{equation}
T_{\rm c}~=~{\Theta_{\rm D} \over 1.45}\exp\left[-{1.04(1+\lambda_{\rm ep}) \over \lambda_{\rm ep} - \mu^*(1+0.62\lambda_{\rm ep})}\right]~,
\label{EqMcMillan1}
\end{equation}
\noindent
which can be inverted to give $\lambda_{\rm ep}$ in terms of $T_{\rm c}$, $\Theta_{\rm D}$ and $\mu^*$ as 

\begin{equation}
\lambda_{\rm ep}~=~{1.04+\mu^*\ln({\Theta_{\rm D} \over 1.45T_{\rm c}})\over (1-0.62\mu^*)\ln({\Theta_{\rm D} \over 1.45T_{\rm c}})-1.04}~.  
\label{EqMcMillan2}
\end{equation}
From the value $\Theta_{\rm D}$~=~539~K obtained above from heat capacity measurements, and using $T_{\rm c}$~=~2.1~K we get $\lambda_{\rm ep} = 0.41$ and~0.50 for $\mu^*$~=~0.10 and 0.15, respectively.  These values of $\lambda_{\rm ep}$ are similar to those found for polycrystalline samples and suggest that OsB$_2$ is a moderate-coupling superconductor.\cite{Singh2007}

\begin{figure}[t]
\includegraphics[width=2.4in, angle=-90]{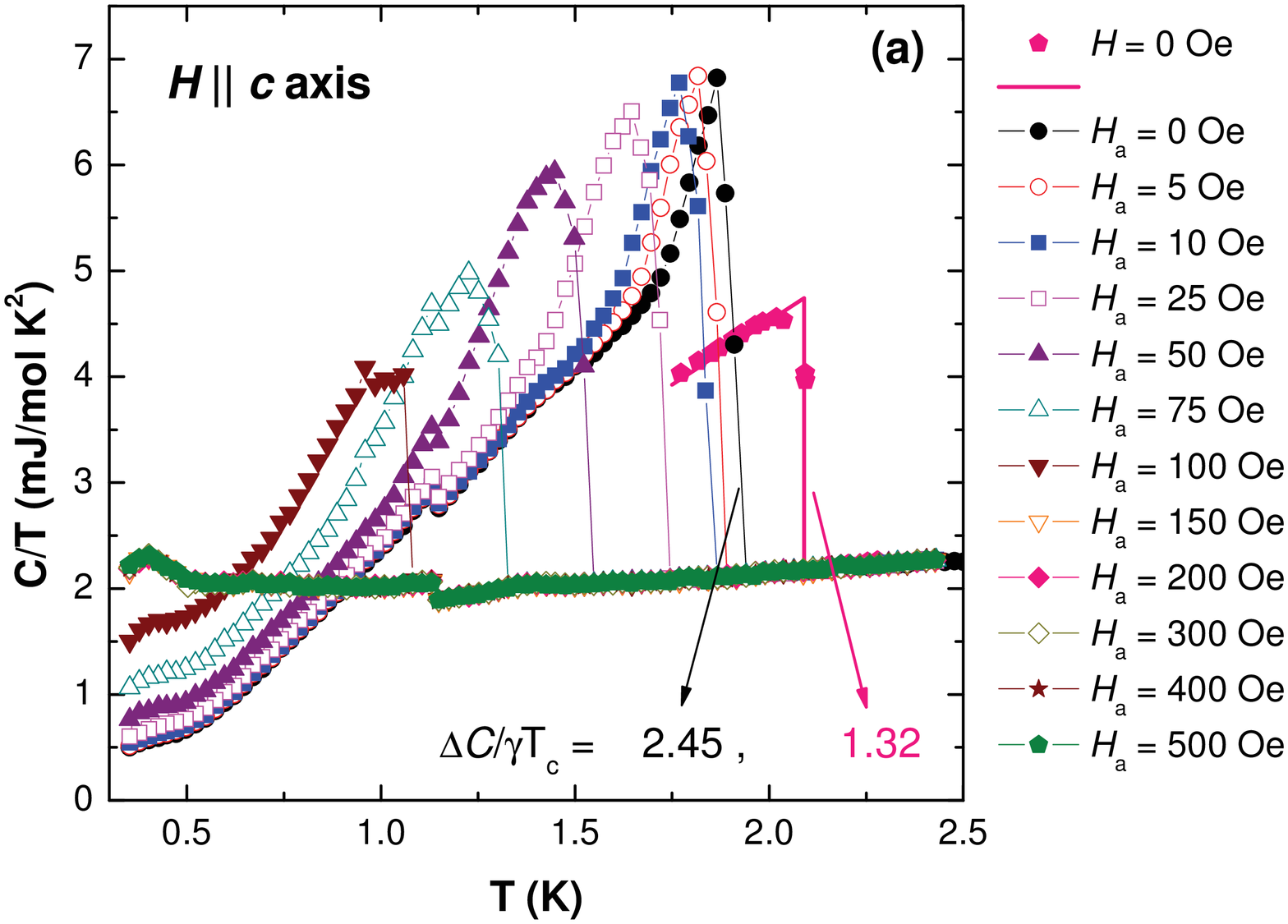}
\includegraphics[width=2.4in, angle=-90]{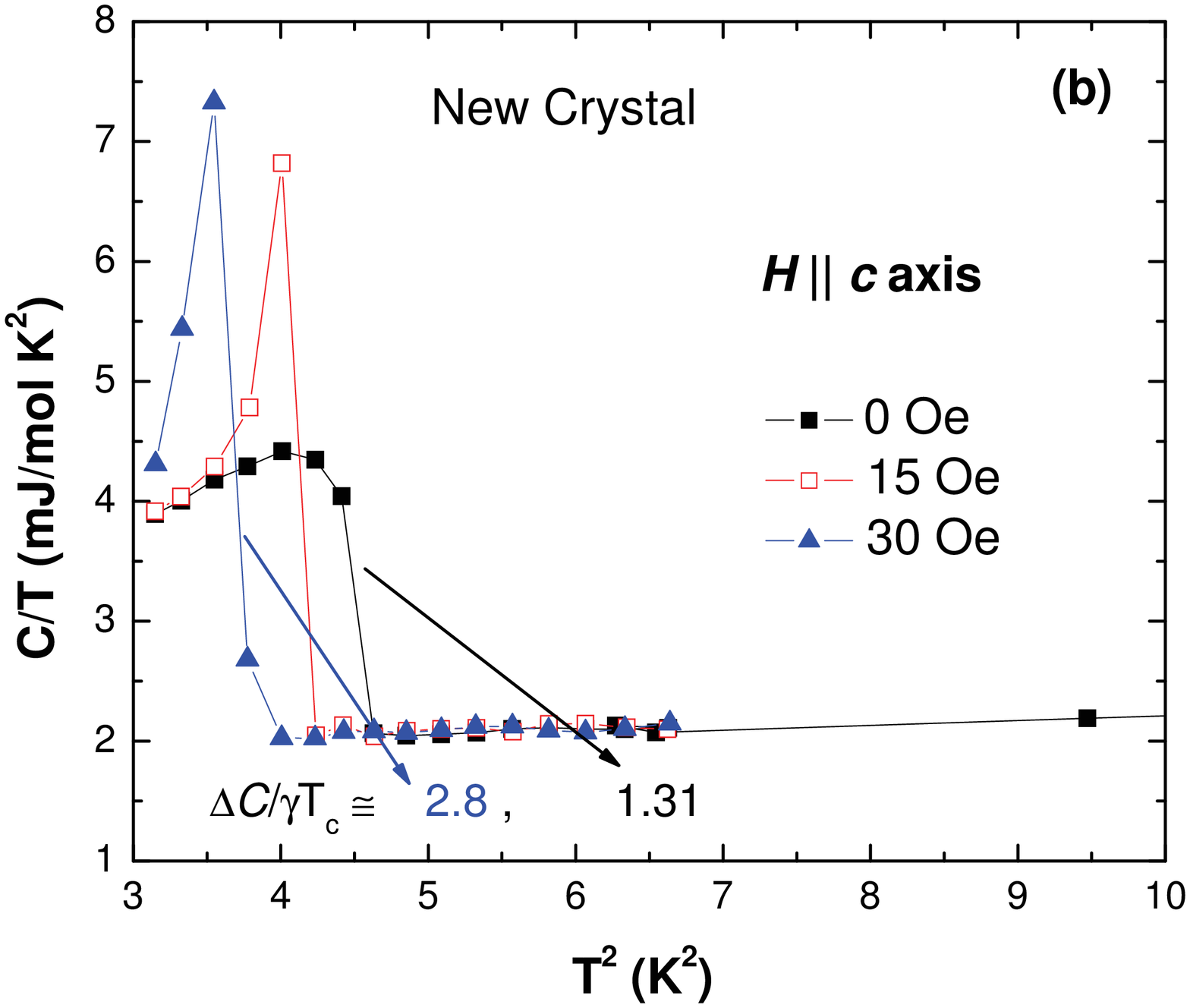}
\caption{(Color online)  (a) Heat capacity $C$ divided by temperature $T$ versus $T$ of a single crystal of OsB$_2$ in various applied magnetic fields $H_{\rm a}$.  A remanent magnetic field $H \approx 25$~Oe was present over the nominal value of the applied field $H_{\rm a}$ listed in the figure legend.  The true $H = 0$~Oe data from Fig.~\ref{Fig(Os,Ru)B2Cp} inset~(b) are also included for comparison.  (b) $C/T$ versus $T^2$ measured in various true magnetic fields $H$ for a different OsB$_2$ crystal.
\label{FigSC-RES-HC-OsB2}}
\end{figure}

The density of states at the Fermi energy $N(\epsilon_{\rm F})$ for both spin directions can be estimated from the values of $\gamma$ and $\lambda_{\rm ep}$ using the relation\cite{Kittel}

\begin{equation}
\gamma = \gamma_0(1+\lambda_{\rm ep})~.
\label{EqDOSHC}
\end{equation}
where

\begin{equation}
\gamma_0 = {\pi^2 k_{\rm B}^2\over 3}N(\epsilon_{\rm F}) = 2.359~N(\epsilon_{\rm F}),
\label{EqDOSHC2}
\end{equation}
$k_{\rm B}$ is Boltzmann's constant and the equality on the right-hand side of Eq.~(\ref{EqDOSHC2}) is for $\gamma_0$ in mJ/mol~K$^2$ and $N(\epsilon_{\rm F})$ in states/(eV f.u.) for both spin directions.  Using the above $\gamma = 1.95$~mJ/(mol~K$^2$), we find $N(\epsilon_{\rm F}) = 0.59$ and 0.55~states/(eV~f.u.) for the above $\lambda_{\rm ep} = 0.41$ and 0.50, respectively.  These values are in excellent agreement with the value from band structure calculations [$N(\epsilon_{\rm F}) = 0.55$~states/(eV~f.u.) for both spin directions].\cite{Hebbachea2006}  This agreement indicates that OsB$_2$ is a weakly correlated electron system, consistent with the observed diamagnetic susceptibility in Fig.~\ref{Figchi_normal}(b) above.  Henceforth we will take the bare density of states to be

\begin{equation}
N(\epsilon_{\rm F}) = 0.55~{\rm states/(eV~f.u.)} 
\label{EqNEFBand}
\end{equation}
for both spin directions, which corresponds to $\mu^* = 0.15$ and $\lambda_{\rm ep} = 0.50$.

To obtain the critical magnetic field versus temperature we have measured $C(T)$ in various $H$.  Figure~\ref{FigSC-RES-HC-OsB2}(a) shows the $C(T)/T$ versus $T$ data between $T$~=~0.3~K and 2.5~K, measured in various applied magnetic fields $H_{\rm a}$.  A remanent field of about 25~Oe was present in addition to the applied magnetic field $H_{\rm a}$ in these measurements.  For comparison the true $H$~=~0~Oe data from Fig.~\ref{Fig(Os,Ru)B2Cp} inset~(b) are also shown.  The superconducting transition seen as an abrupt jump in $C(T)/T$ is suppressed to lower $T$ with increasing $H_{\rm a}$ as expected.  However, the magnitude $\Delta C/\gamma T_{\rm c}$ of the anomaly at $T_{\rm c}$ is initially larger than that observed in zero magnetic field.  As shown by the arrows in Fig.~\ref{FigSC-RES-HC-OsB2}(a), $\Delta C/\gamma T_{\rm c}$ increases from 1.32 for $H$~=~0~Oe to 2.45 for $H_{\rm a} = 0$~Oe (which is close to $H = 25$~Oe) suggesting a divergent nature of $C$ at $T_{\rm c}$ in an applied magnetic field.  

The superconducting anomaly moves to lower $T$ with increasing $H$ and is not observed above $H_a = 100$~Oe ($H \approx 125$~Oe).  The step in all data at about $T = 1.15$~K arises from a problem in the measurement and is not intrinsic to the sample.  

To further study the enhanced $\Delta C$ anomaly in low fields we measured $C(T)$ for another single crystal in various (true) $H$.  The data are plotted as $C(T)/T$ versus $T^2$ in Fig.~\ref{FigSC-RES-HC-OsB2}(b).  We again observe that the anomaly at the superconducting transition becomes first order-like in a finite field showing that this feature is intrinsic to single crystalline OsB$_2$.

\begin{figure}[t]
\includegraphics[width=2.75in,angle=-90]{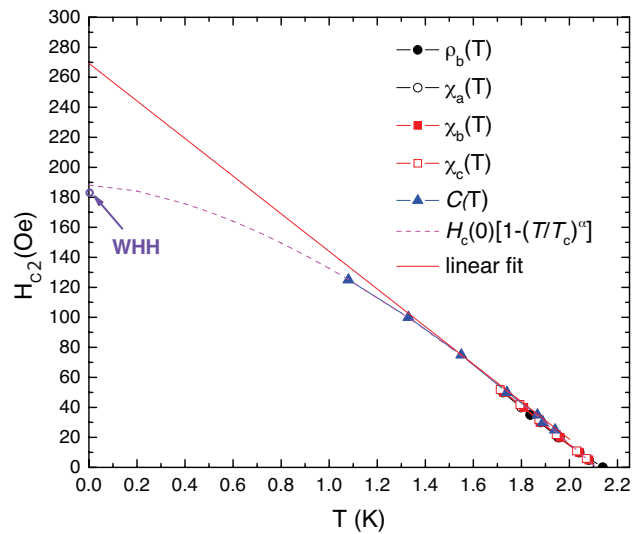}
\caption{(Color online) Upper critical magnetic field $H_{\rm c2}$ versus temperature $T$ extracted from different types of measurements, as indicated.  The straight line is a linear fit to the data near $T_{\rm c}$.  The dashed curve is a fit by the expression $H_{\rm c2}(T) = H_{\rm c2}(0)[1-({T\over T_{\rm c}})^{\alpha}]$.  The circle at $T = 0$~K labeled ``WHH'' is the estimate of $H_{\rm c2}(T = 0)$ using the WHH formula in the clean limit (see text).  
\label{Figcrit_field}}
\end{figure}

This $\Delta C(H)$ behavior is similar to that recently observed for $T_2$Ga$_9$ ($T = $ Rh and Ir),\cite{Shibayama2007,Wakui2009} and for single crystals of ZrB$_{12}$,\cite{Wang2005} where it was suggested that the Type-I superconductivity in these materials led to the superconducting transition in a finite magnetic field to be first order-like, resulting in a divergent $\Delta C(H)$ at $T_{\rm c}$.  A similar divergent $\Delta C(H)$ at $T_{\rm c}$ was observed 75 years ago for the Type-I superconductor thallium.\cite{Smith1935}  The $C(T,H)$ behavior observed for single crystal OsB$_2$ in Figs.~\ref{FigSC-RES-HC-OsB2}(a) and~(b) is similar to that observed for the materials mentioned above and might suggest that OsB$_2$ is a Type-I superconductor.  However, our estimates of the Ginzburg-Landau parameter $\kappa = 2(1)$ below indicate that OsB$_2$ is a small-$\kappa$ Type-II superconductor.  The unusual features in the $C(T,H)$ for OsB$_2$ are therefore not understood at present but might be related to the multi-gap nature of the superconductivity.

\subsection{Upper Critical Magnetic Field $H_{\rm c2}(T)$}  
\label{sec:RES-SC-criticalfield-superfluiddensity}
The $H_{\rm c2}(T)$ data obtained from all of the above measurements are plotted in Fig.~\ref{Figcrit_field}.  In the temperature range of the SQUID magnetometer measurements (1.7~K to 2.4~K) all the data match well and the temperature dependence of $H_{\rm c2}$ is linear (solid curve extrapolated to $T = 0$~K in Fig.~\ref{Figcrit_field}) with the slope ${dH_{\rm c} \over dT} = -125$~Oe/K\@.  This linear slope can be used to get an estimate of the $T = 0$~K upper critical field using the WHH formula for the clean limit $H_{\rm c2}(0) = -0.693\, T_{\rm c} ({dH_{\rm c} \over dT}|_{T_{\rm c}})$.\cite{WHH1966}  Using the above value of ${dH_{\rm c} \over dT}|_{T_{\rm c}} = -125$~Oe/K and $T_{\rm c} = 2.10$~K we get $H_{\rm c2}(0) = 182$~Oe.  

The $H_{\rm c2}(T)$ data at the lower temperatures $T\sim 1$~K in Fig.~\ref{Figcrit_field} show a deviation from linearity with a negative curvature.  To obtain another estimate of $H_{\rm c2}(0)$, the $H_{\rm c2}(T)$ data in the whole $T$ range were fitted by the empirical power law expression $H_{\rm c2}(T) = H_{\rm c2}(0)\left[1-({T\over T_{\rm c}})^{\alpha}\right]$ with $H_{\rm c2}(0)$ and $\alpha$ as fitting parameters and with fixed $T_{\rm c} = 2.15$~K\@.  The fit shown as the dashed curve in Fig.~\ref{Figcrit_field} gave the values $H_{\rm c2}(0) = 188(2)$~Oe and $\alpha = 1.55(3)$.  This estimate of $H_{\rm c2}(0)$ is close to the value of 182~Oe obtained above using the WHH formula.  These two fits together yield our final value $H_{\rm c2}(0) = 186(4)$~Oe.  From Fig.~\ref{Figcrit_field} it can also be seen that there is negligible anisotropy in the measured $H_{\rm c2}(T)$ from $T = 1.7$ to 2.4~K\@.  

For a Type-II superconductor near $T_{\rm c}$, the superconducting coherence length $\xi$ can be estimated from the measured $H_{\rm c2}$ using the Ginzburg-Landau relation\cite{Tinkham} 

\begin{equation}
H_{\rm c2} = \frac{\phi_0}{2\pi\xi^2},  
\label{Eqkappa2}
\end{equation}
where $\phi_0 = hc/2e = 2.068 \times 10^{-7}$~G~cm$^2$ is the flux quantum.  We obtain an estimate of $\xi$ using instead the zero-temperature value $H_{\rm c2}(T=0) =  186(4)$~Oe arrived at above to obtain $\xi(T = 0) = 0.133(2)~\mu$m.

\subsection{\label{SecSFD} Superfluid Density}
The measured magnetic penetration depth $\lambda(T)$ in the superconducting state is related to the so-called London pentration depth $\lambda_{\rm L}(T)$ by\cite{Tinkham}

\begin{equation}
\lambda(T) \approx \lambda_{\rm L}(T)\sqrt{1 + \frac{\xi_0(T)}{\ell(T)}},
\label{EqLambda}
\end{equation}
where 

\begin{equation}
\xi_0 = \frac{\hbar v_{\rm F}}{\pi \Delta(0)}
\label{Eqxi0}
\end{equation}
is the BCS coherence length, $\ell$ is the quasiparticle mean free path and $v_{\rm F}$ is the Fermi velocity.  Including the influence of $\ell$ gives the modified coherence length $\xi$ as\cite{Tinkham}

\begin{equation}
\frac{1}{\xi} = \frac{1}{\xi_0} + \frac{1}{\ell}.
\label{Eqxi}
\end{equation}
The limit $\xi/\ell\to 0$ is called the clean limit and the opposite limit the dirty limit.  

The superfluid density $\rho_{\rm s}(T)$ is related to $\lambda(T)$ by\cite{Tinkham}

\begin{equation}
\rho_{\rm s}(T) = \frac{m^* c^2}{4\pi e^2\lambda^2(T)},
\label{Eqnslambda}
\end{equation}
where $m^*$ is the effective mass of the individual quasiparticles, $c$ is the speed of light in vacuum, $e$ is the elementary charge and $\rho_{\rm s}$ is the density of quasiparticles that have condensed into the superconducting state, not the density of Cooper pairs which is a factor of two smaller. 
The normalized ratio of $\rho_{\rm s}(T)$ to $\rho_{\rm s}(0)$ is simply

\begin{equation}
\frac{\rho_{\rm s}(T)}{\rho_{\rm s}(0)} = \frac{\lambda^2(0)}{\lambda^2(T)}.
\label{Eqrhoratio}
\end{equation}

We now estimate whether OsB$_2$ is in the clean or dirty limit or somewhere in between, by estimating the ratio $\xi(0)/\ell$.  The value of $\xi(0)$ was derived in the preceding section.  We will estimate the mean-free-path $\ell$ using the measured resistivity at low temperatures and the $N(\epsilon_{\rm F})$ in Eq.~(\ref{EqNEFBand}).  First, the conductivity $\sigma$ is written as\cite{Kittel}

\begin{equation}
\sigma = \frac{ne^2\tau}{m^*}
\label{Eqsigms}
\end{equation}
where $n$ is the conduction carrier density and $\tau$ is the mean-free scattering time of the current carriers.  We then express $\tau = v_{\rm F}/\ell$, and from Eq.~(\ref{Eqsigms}) we get

\begin{equation}
\ell = \frac{\hbar}{e^2}\frac{m^*v_{\rm F}\sigma}{\hbar n},
\label{Eqell2}
\end{equation}
where in SI units the first term on the right is $\hbar/e^2 = 4108~\Omega$.  Next we write both $v_{\rm F}$ and $n$ in terms of the known $N(\epsilon_{\rm F})$ and then substitute these expressions into Eq.~(\ref{Eqell2}).

\begin{figure}[t]
\includegraphics[width=3in]{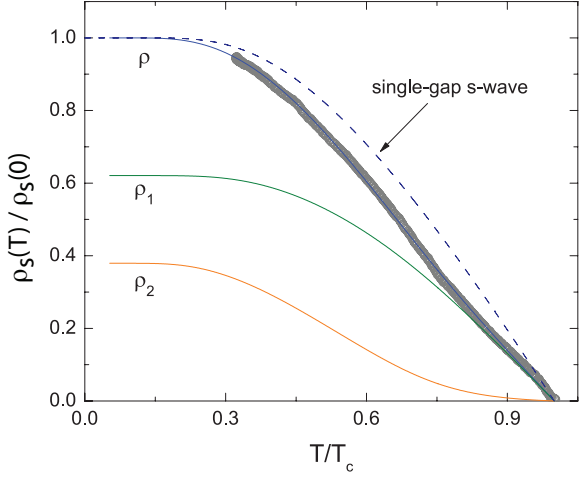}
\caption{(Color online) The normalized superfluid density $\rho_{\rm s}(T)/\rho_{\rm s}(0)$ versus reduced temperature $T/T_{\rm c}$.  The solid curve through the data is a fit by the two-gap $\gamma$ model for superconductivity.  The curves $\rho_1$ and $\rho_2$ are the individual contributions from the two gaps. 
\label{Fig-rho}}
\end{figure}

The (average) Fermi velocity has not been reported from band calculations.  Therefore we calculate both  $v_{\rm F}$ and $n$ from $N(\epsilon_{\rm F})$ by assuming a three-dimensional single-band model with a spherical Fermi surface, yielding\cite{Kittel}

\begin{equation}
v_{\rm F} = \frac{\pi^2\hbar^3}{2m_{\rm e}^2({m^*}/m_{\rm e})^2}{\cal D}(\epsilon_{\rm F}),
\label{EqvF1}
\end{equation}

\begin{equation}
n = \frac{\pi^4}{3}\left[\frac{\hbar^2}{2m^*}{\cal D}(\epsilon_{\rm F})\right]^3,
\label{Eqn}
\end{equation}
where ${\cal D}(\epsilon_{\rm F})$ is the density of states at the Fermi energy in units of states/(erg cm$^3$) for both spin directions.  Substituting Eqs.~(\ref{EqvF1}) and~(\ref{Eqn}) into~(\ref{Eqell2}), and using $\rho = 1/\sigma$, gives

\begin{equation}
\ell = \frac{\hbar}{e^2}\frac{3(m^*/m_{\rm e})^2}{\pi^2\left[\frac{\hbar^2}{2m_{\rm e}}{\cal D}(\epsilon_{\rm F})\right]^2 \rho},
\label{EqellDOS}
\end{equation}
where $m_{\rm e}$ is the free-electron mass.   The expression converting ${\cal D}(\epsilon_{\rm F})$ in units of ${\rm states/(erg~cm^3)}$ to the conventional units of states/(eV~f.u.) for both spin directons appropriate to the above definition of $N(\epsilon_{\rm F})$ is

\begin{equation}
{\cal D}(\epsilon_{\rm F}) = N(\epsilon_{\rm F})\left[\frac{1}{\rm eV~f.u.}\right]\left({\rm \frac{1~eV}{1.6022\times10^{-12}~erg}}\right)\frac{N_{\rm A}}{V_{\rm M}},
\label{EqConvertDN}
\end{equation}
where $N_{\rm A}$ is Avogadro's number and $V_{\rm M}$ is the molar volume.  Substituting Eq.~(\ref{EqConvertDN}) into~(\ref{EqellDOS}) and putting in the values of the constants gives

\begin{equation}
\ell = 2.372\times 10^{-14} \frac{(m^*/m_{\rm e})^2V_{\rm M}^2}{N^2(\epsilon_{\rm F})\rho},
\label{Eqellfinal}
\end{equation}

\begin{equation}
v_{\rm F} = 2.622 \times 10^9 \frac{N(\epsilon_{\rm F})}{(m^*/m_{\rm e})^2 V_{\rm M}}
\label{EqvF2}
\end{equation}
where $\ell$ is in cm, $v_{\rm F}$ is in cm/s, $N(\epsilon_{\rm F})$ is in states/(eV~f.u.) for both spin directions, $V_{\rm M}$ is in cm$^3$/mol and $\rho$ is in $\Omega$~cm.

Inserting $V_{\rm M} = 16.47$~cm$^3$/mol from Table~\ref{CrystalData}, $m^*/m_{\rm e} =1$ (see Sec.~\ref{SdH} below), $N(\epsilon_{\rm F}) = 0.55$~states/(eV f.u.) for both spin directions from our heat capacity data above, and $\rho = 1.55 \times 10^{-6}~\Omega$~cm at 2.25~K from Fig.~\ref{Fig-OsB2RES} into Eq.~(\ref{Eqellfinal}) gives $\ell = 0.137~\mu$m at 2.25~K\@.  Then using $\xi(0) = 0.133$~$\mu$m from above gives $\xi(0)/\ell(0) = 0.97$.  Therefore OsB$_2$ is in neither the clean limit nor the dirty limit, but in between.  Irrespective of this difficulty, we will assume the clean limit in order to be able to carry out calculations for comparison with our measured penetration depth data.  From Eq.~(\ref{EqvF2}) we also obtain $v_{\rm F}= 8.75\times10^7$~cm/s.

\begin{figure}[t]
\includegraphics[width=3in]{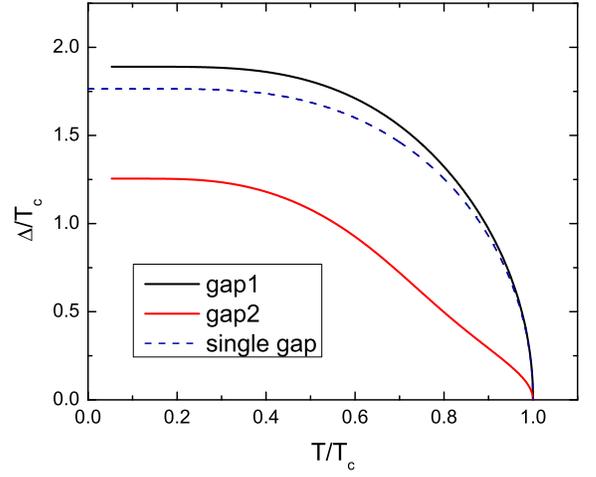}
\caption{(Color online)  Temperature $T$ dependence of the two gaps $\Delta_1$ (top black solid curve) and $\Delta_2$ (bottom red solid curve). Also shown by the blue dashed curve is the BCS prediction for a single gap.
\label{Fig-Delta}}
\end{figure}

The $\rho_{\rm s}(T)/\rho_{\rm s}(0)$ versus $T/T_{\rm c}$ calculated from the $\lambda(T)$ data using Eq.~(\ref{Eqrhoratio}) is shown in Fig.~\ref{Fig-rho}.  The dashed curve is the $T$ dependence of $\rho_{\rm s}$ expected for a BCS single-gap $s$-wave superconductor.  It is clear that our $\rho_{\rm s}(T)$ shows marked deviations from the single-gap BCS curve.  This is consistent with our previous observations for polycrystalline samples.\cite{Singh2007}  The solid curve through the data is a fit by a two-gap $\gamma$ model.\cite{Kogan2009}  From the fit we obtained $\lambda(T\to 0) = 0.300~\mu$m.

The partial superfluid densities $\rho_1(T)$ and $\rho_2(T)$ from the fit to the two-gap model are shown as solid curves in Fig.~\ref{Fig-rho}.  The $T$ dependences of the two gaps are shown in Fig.~\ref{Fig-Delta} plotted as normalized gaps $\Delta/k_{\rm B}T_{\rm c}$ versus the reduced temperature $T/T_{\rm c}$.  For comparison, the $T$ dependence of a single $s$-wave BCS gap is shown as the dashed curve.  The $T = 0$ value of the two gaps are $\Delta_1(0) = 1.88 k_{\rm B}T_{\rm c}$ and $\Delta_2(0) = 1.25k_{\rm B}T_{\rm c}$, respectively.  The ratios of these two gaps to the single BCS gap value $\Delta_{\rm BCS}(0) = 1.76 k_{\rm B}T_{\rm c}$ are $\Delta_1(0)/\Delta_{\rm BCS}(0) = 1.07$ and $\Delta_2(0)/\Delta_{\rm BCS}(0) = 0.71$.  The values of these two gaps agree by construction with the theorem that in a two-gap superconductor, one of the gaps will always be larger than the BCS gap, whereas the second will always be smaller.\cite{Kresin1990}  This constraint is a built-in result of the self-consistent solution to the two-gap $\gamma$ model.

\subsection{Additional Superconducting Parameters}
The zero-temperature thermodynamic critical field $H_{\rm c}(0)$ of a superconductor is related to the zero-temperature superconducting gap $\Delta(0)$ in a single-gap BCS model by the expression\cite{Tinkham}

\begin{equation}
{H_{\rm c}(0)^2\over 8\pi} = {{\cal D}(\epsilon_{\rm F})\Delta(0)^2\over 4}~,
\label{Eqcondensationenergy}
\end{equation}
where, as above, ${\cal D}(\epsilon_{\rm F})$ is the density of states at the Fermi energy for both spin directions in units of states/(erg~cm$^3$).  We use this expression as an approximation to our two-gap model to obtain a value of $H_{\rm c}(0)$.  Using the density of states value $N(\epsilon_{\rm F}) = 0.55$ states/eV~f.u.\ for both spin directions from the above heat capacity measurements and Eq.~(\ref{EqConvertDN}) gives ${\cal D}(\epsilon_{\rm F}) = 1.26\times10^{34}$~states/(erg~cm$^3$).  Using the larger gap found from fitting the penetration depth data and $\Delta(0)/k_{\rm B}T_{\rm c} = 1.88$ which gives $\Delta(0) = 5.45 \times 10^{-16}$~erg, Eq.~(\ref{Eqcondensationenergy}) yields $H_{\rm c}(0) = 153$~Oe.  We can now derive the Ginzburg-Landau parameter $\kappa$ using the above $H_{\rm c2}(0) = 186$~Oe via\cite{Tinkham}

\begin{equation}
\kappa = \frac{H_{\rm c2}}{\sqrt{2}H_{\rm c}}= 0.86.
\end{equation}
This value is marginally on the Type-II side of the value $\kappa = 1/\sqrt{2} \approx 0.707$ separating Type-I from Type-II superconductivity, thus justifying the above notation of the measured critical field as being teh upper critical field $H_{\rm c2}$ instead of the thermodynamic critical field $H_{\rm c}$.

Another estimate of $\kappa$ can be obtained using the relation\cite{Parks}

\begin{equation}
\kappa(T) = {\kappa(0)\over [1+(T/T_{\rm c})^2]} = 2^{1/2}{2\pi H_{\rm c}(0) \lambda(0)^2\over \phi_0 [1+(T/T_{\rm c})^2]}~,
\label{Eqkappa0}
\end{equation}
where $\kappa(0)$, $H_{\rm c}(0)$, and $\lambda(0)$ are the $T = 0$ values of the Ginzburg-Landau parameter, the thermodynamic critical field, and penetration depth respectively.  With the value $H_{\rm c}(0) = 153$~Oe obtained above and the value $\lambda(0) = 0.300$~$\mu {\rm m}$, we get $\kappa(0) = 6.0$ and $\kappa(T_{\rm c}) = 3.0$.

Two more estimates of $\kappa(T_{\rm c})$ can be made using the relations\cite{Tinkham}

\begin{equation}
\kappa(T_{\rm c}) = 0.96{\lambda_{\rm L}(0)\over \xi(0)},\ \ \ \ ({\rm clean~limit})
\label{Eqkappa1}
\end{equation} 

\begin{equation}
\kappa(T_{\rm c}) = 0.715{\lambda_{\rm L}(0)\over \ell}.\ \ \ \ \ \ ({\rm dirty~limit})
\label{Eqkappa3}
\end{equation} 
Using $\lambda(0) = 0.300$~$\mu {\rm m}$, $\xi(0) = 0.133$~$\mu$m, $\ell(T>T_{\rm c}) = 0.137~\mu$m and Eq.~(\ref{EqLambda}), one obtains $\kappa(T_{\rm c}) = 2.17$ and~0.78 from Eqs.~(\ref{Eqkappa1}) and~(\ref{Eqkappa3}), respectively.

The above four estimates of $\kappa(T_{\rm c})$ are all greater than $1/\sqrt{2}$ and therefore all indicate that single crystalline OsB$_2$ is a small-$\kappa$ Type-II superconductor with $\kappa(T_{\rm c}) = 2(1)$.

\begin{figure}[t]
\includegraphics[width=3.3in]{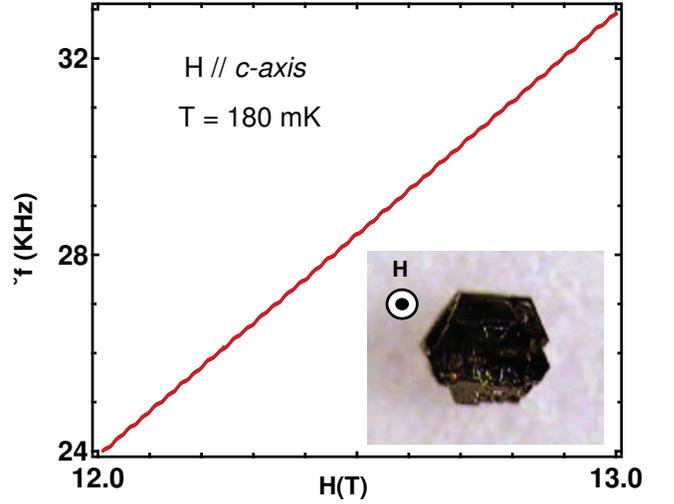}
\caption{(Color online) The change $\Delta f = f(H) - f(0)$ in the tunnel diode oscillator (TDO) frequency $f$ versus magnetic field $H$ measured at $T = 180$~mK with $H$ applied along the $c$ axis.  The inset shows an image of the crystal used for the measurement.  The $c$ axis is out of the plane of the image.
\label{Fig-sdh}}
\end{figure}

\begin{figure}[t]
\includegraphics[width=2.65in,angle=-90]{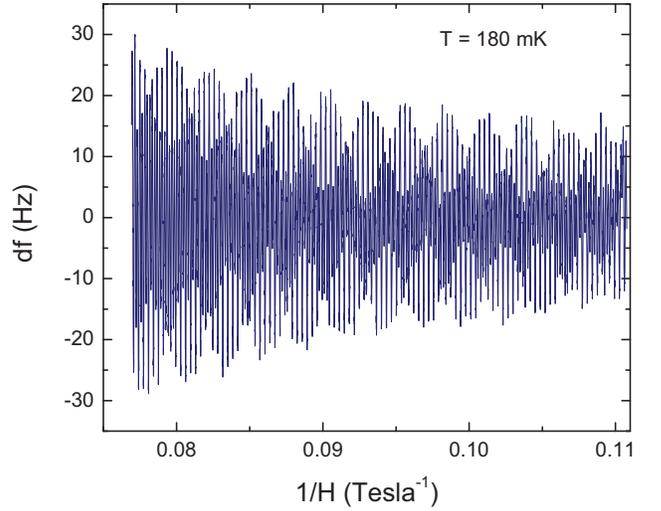}
\caption{(Color online) The oscillating part of the TDO frequency $df = \Delta f -$ smooth background, versus the reciprocal of the magnetic field $1/H$ for $H$ applied along the $c$~axis.    
\label{Fig-sdh2}}
\end{figure}

\subsection{\label{SdH} Shubnikov-de Haas (SdH) Oscillations}
Shubnikov-de Haas (SdH) oscillations in $\rho(H)$ were observed as oscillations in the skin depth, which in turn were obtained from the oscillation frequency shift versus $H$ of a tunnel diode oscillator (TDO) in which the sample is placed inside the inductor of the $LC$ circuit.  Oscillations were observed for $T = 0.12$--$3.3$~K in magnetic fields up to $H = 14$~T\@.  Since $H_{\rm c2}(0)$ = 186~Oe from Sec.~\ref{sec:RES-SC-criticalfield-superfluiddensity}, such fields quench the superconductivity and the measurements are therefore in the normal state.  The inset of Fig.~\ref{Fig-sdh} shows an image of the crystal and the direction of the applied field $H || c$~axis where the $c$~axis points out of the plane of the figure.  

The shift $\Delta f = f(H) - f(0)$ in the TDO frequency versus $H$ applied along the $c$ axis measured at $T = 0.18$~K is shown versus $H$ in Fig.~\ref{Fig-sdh}.  Small oscillations can be seen riding on a smooth background.  This $H$-dependent background is due to the tunnel diode circuit that is partially exposed to the applied field.  The oscillations are more clearly visible when a smooth background is subtracted from $\Delta f(H)$ using a non-oscillating piecewise cubic hermite interpolating polynomial algorithm in Matlab.   Figure~\ref{Fig-sdh2} shows the resulting oscillating part of the TDO frequency shift $df$ versus the inverse magnetic field $1/H$ where $df$ is the frequncy shift after the background subtraction.

\begin{table}
\caption{\label{SdHFreqs} Experimental (expt) and theoretical\cite{Hebbache-FS} (thy) Shubnikov-de Haas frequencies $F$ for single crystal OsB$_2$ with the magnetic field applied along the $c$~axis and in the $ab$~plane, where $n$ is the harmonic number.  Also included are values for the effective mass $m^*/m_{\rm e}$ and the electron-phonon coupling constant $\lambda_{\rm ep}$ of the electrons is specific orbits. For the data with $H\parallel ab$~plane, the field was oriented about 15$^\circ$ from the $b$~axis, and hence the measured frequencies are not in general equal to the theoretical values for $H \parallel b$.}

\begin{ruledtabular}
\begin{tabular}{lcccccc}
$F$ & $n$ &  $F_{\rm expt}$ & $F_{\rm thy}$ & $\frac{m^*}{m_{\rm e}}$ & $\frac{m^*}{m_{\rm e}}$ & $\lambda_{\rm ep}$\\
&& (T) & (T) & expt & thy & expt\\ \hline
${\bf H \parallel c}$\\
$F_1$ & 1 & 2767 & 3023 & 1.03(4) & 1.05 & $\approx 0$\\
$F_1$ & 2 & 5528 & --- & --- & ---\\
$F_2$ & 1 & 5905 & 5902 & 0.81(3) & 0.50 & 0.63(5)\\
$F_2$ & 2 & 11\,806 & --- & ---& ---\\
$F_3$ &  & 660 &  \\ \hline
${\bf H \parallel ab}$\\
$F_1$ &  & 745\\
$F_2$ & & 932 \\
$F_3$ &  & 2983 & $\sim$ 2177 \\
$F_4$ &  & 3812 & $\sim$ 3265 & 0.87 & 0.95 & $\approx 0$\\
$F_5$ &  & 3957 & $\sim$ 3888 & 1.13 & 0.92 & 0.23\\
$F_6$ &  & 5138 & 5115 & 0.92 & 0.62 & 0.48\\
$F_7$ & 1? & 5697 & $\sim$ 5291 \\
$F_8$ &  & 6153 & 6189 & 0.96 & 0.45 & 1.13\\
$F_9$ & 2? & 11\,311 \\
\end{tabular}
\end{ruledtabular}
\end{table}

\begin{figure}[t]
\includegraphics[width=\columnwidth]{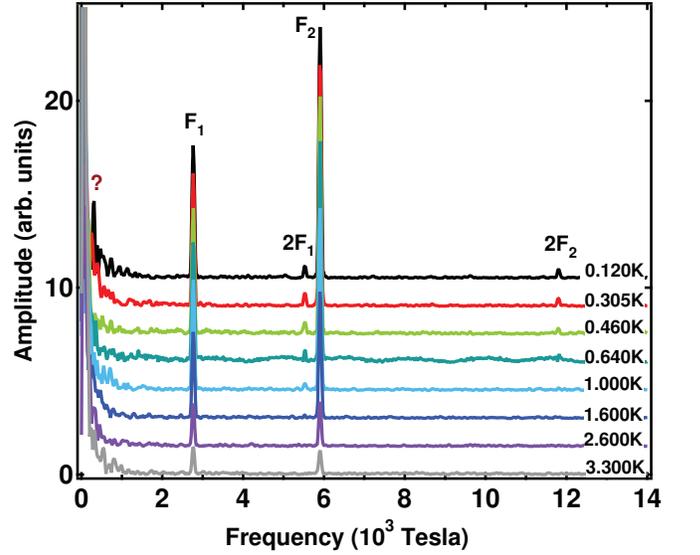}
\caption{(Color online) Power spectra of Shubnikov-de Haas oscillations obtained at the indicated temperatures from Fourier transformation of the quantum SdH oscillation data such as in Fig.~\ref{Fig-sdh2} at the indicated temperatures. The spectra are shifted vertically by arbitrary amounts for clarity of presentation. 
\label{Fig-power}}
\end{figure}

To get the frequencies of the SdH oscillations at each $T$, a power spectrum was obtained by taking a Fourier transformation of the oscillation data such as in Fig.~\ref{Fig-sdh2}.  The resulting power spectra obtained for the measurements at $T = 0.12$--$3.3$~K are shown in Fig.~\ref{Fig-power}.  The data reveal two clear fundamental frequencies $F_1 = 2767$~T and $F_2 = 5905$~T and possibly a third $F_3 = 660$~T as marked in the plot in Fig.~\ref{Fig-power}, although the intensity of the line at $F_3$ is much weaker than the intensities of the prominent sharp lines at $F_1$ and $F_2$.  We also observe the second harmonics for $F_1$ and $F_2$ but none for $F_3$, as shown in Fig.~\ref{Fig-power} and listed in Table~\ref{SdHFreqs}.

Recent first principles calculations of the Fermi surface (FS) showed three bands at the Fermi level, consisting of two nested deformed ellipsoidal surfaces (first and second bands) and a a corrugated tubular surface (third band) along the $b$ axis.\cite{Hebbache-FS}  For a magnetic field applied along the $c$ axis, the two closed electronic orbits which give rise to SdH oscillations are the cross-sectional areas of the two deformed ellipsoids normal to the applied field.  The theoretically predicted frequencies of oscillations are 3023 and 5902~T.\cite{Hebbache-FS}  These values are in reasonable agreement with the experimentally observed frequencies $F_1 = 2767$ and $F_2 = 5905$~T of the SdH oscillations for OsB$_2$ for measurements with $H || c$ axis.  Since the frequencies of the SdH oscillations are inversely proportional to the area of the respective electronic orbits, we can assign the $F_1$ oscillations as coming from the smaller inner ellipsoid while the $F_2$ oscillations can be assigned to the outer ellipsoid.  For the third tubular band, there are no closed orbits for $H || c$ axis.  The origin of the third frequency $F_3$ in Fig.~\ref{Fig-power} is therefore not understood at present.   

\begin{figure}[t]
\includegraphics[width=2.5in]{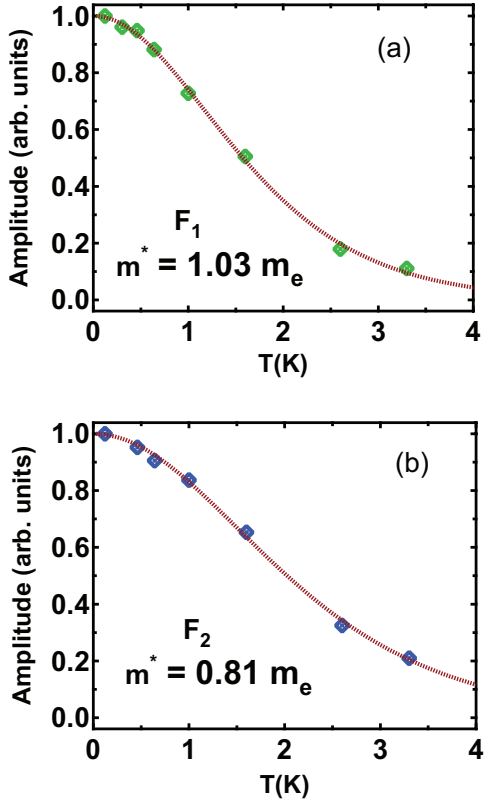}
\caption{(Color online)  (a) and (b) Fits to the amplitude versus $T$ data for the peaks $F_1$ and $F_2$ respectively, by Eq.~(\ref{Eq-LK(T)}).  The values of the respective effective masses $m^*$ for the two Fermi surface extrema are indicated in the figures. 
\label{Fig-power(T)}}
\end{figure}

The $T$ dependences of the amplitudes $A(T)$ of oscillation in Fig.~\ref{Fig-power} can be used to estimate the respective effective masses for the bands responsible for the oscillations.  The normalized $A(T)$ is given by the Lifshitz-Kosevich formula\cite{Lifshitz1956, Shoenberg1984}

\begin{equation}
A(T) = {X\over {\rm sinh}X}~,~~ {\rm where}~~ X = \pi^2\left({\frac{m^*}{m_{\rm e}}}\right)\left(\frac{k_{\rm B}T}{\mu_{\rm B} B}\right),
\label{Eq-LK(T)}
\end{equation}   
and $B \approx H$ is the magnetic induction.  The dimensionless variable $X$ is proportional to the product of the effective mass and the ratio of the thermal to magnetic energies.  Therefore, by fitting the $T$-dependent amplitudes of the peaks in the power spectra in Fig.~\ref{Fig-power} by Eq.~(\ref{Eq-LK(T)}) one can obtain the $m^*$ values for the Fermi surface electrons responsible for the respective oscillations.   Figures~\ref{Fig-power(T)}(a) and (b) show $A(T)$ for the $F_1$ and $F_2$ peaks, respectively.  The fits by Eq.~(\ref{Eq-LK(T)}) are also shown as solid curves through the data.  We obtain $m_1^* = 1.03(4) m_{\rm e}$ for $F_1$ and $m_2^* = 0.81(3) \,m_{\rm e}$ for $F_2$.  The band masses predicted by theory for these two orbits are $m_{\rm b1} = 1.05 m_{\rm e}$ and $m_{\rm b2} = 0.50 m_{\rm e}$, respectively.\cite{Hebbache-FS}  The electron-phonon coupling constant $\lambda_{\rm ep}$ can be estimated by using the expression $m^* = (1+\lambda_{\rm ep}) m_{\rm b}$.  Using the above values of $m^*$ and $m_{\rm b}$ we obtain $\lambda_{\rm ep} \approx 0$ for $F_1$ and $\lambda_{\rm ep} = 0.63(5)$ for $F_2$.

\begin{figure}[t]
\includegraphics[width=3.3in]{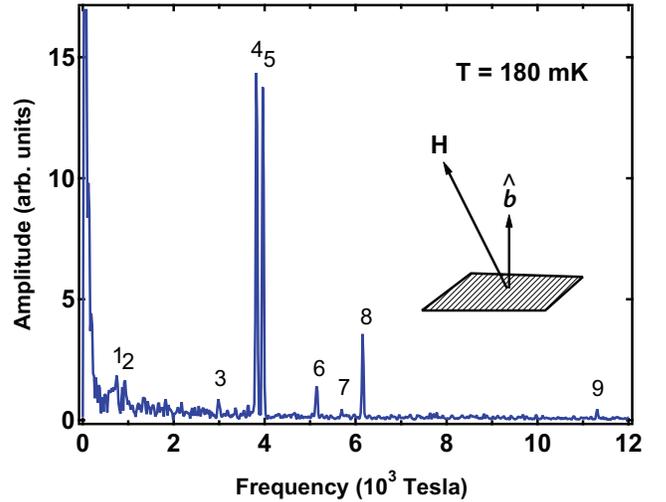}
\caption{(Color online) Power spectra of Shubnikov-de Haas oscillations obtained at temperatures $T = 180$~mK for $H$ about 15$^\circ$ away from the $b$-axis in the $ab$-plane (see inset), obtained from Fourier transformation of  quantum oscillation data (not shown) similar to those in Fig.~\ref{Fig-sdh2}.  The indices $i$ for the quantum SdH oscillation frequencies $F_i$ are as indicated.  These frequencies are listed in Table~\ref{SdHFreqs}.
\label{Fig-power2}}
\end{figure}

Figure~\ref{Fig-power2} shows the power spectra for quantum oscillations measurements performed with $H$ applied perpendicular to the $c$ axis.  Due to the shape of our crystal we were only able to perform measurements with $H$ tilted about $15$ degrees away from the $b$ axis as shown in the inset of Fig.~\ref{Fig-power2}.  The power spectra reveal nine different frequencies which are labeled in Fig.~\ref{Fig-power2} and their values are given in Table~\ref{SdHFreqs}.  

For $H || b$ axis, theory predicts at least six frequencies for quantum oscillations, two coming from the deformed ellipsoidal Fermi surfaces, and four from different closed orbits on the tubular Fermi surface as shown in Fig.~3 in Ref.~\onlinecite{Hebbache-FS}.  For the two deformed ellipsoidal FS sheets, it was predicted that the frequencies of oscillation would vary linearly with angle as the field $H$ is moved from $H~||$ $b$ axis toward $H~ ||$ $c$ axis.\cite{Hebbache-FS}  Thus, for the first deformed ellipsoid the frequency of oscillation should change at a rate of about $-25$~T/degree and for the second deformed ellipsoid it should change at a rate of about $-3.8$~T/degree as one moves away from $H \parallel b$~axis.\cite{Hebbache-FS}  The predicted values of the frequencies of oscillation for $H \parallel b$~axis are 5490~T and 6246~T, respectively, for the two deformed ellipsoids.  If we use the above rates of change of the frequencies and the fact that we measured with $H$ about 15 degrees away from the $b$ axis, then the expected frequencies are 5115~T and 6189~T, respectively, as listed in Table~\ref{SdHFreqs}.  These values are close to the observed frequencies $F_6$ and $F_8$, respectively.  Therefore, we can assign the frequencies $F_6$ and $F_8$ as coming from the two deformed ellipsoidal FS sheets.  The assignment of the other frequencies in Table~\ref{SdHFreqs} is difficult since the angular dependence of frequencies arising from the tubular FS sheet is not known.  We note that frequencies $F_1$ and $F_2$ in Fig.~\ref{Fig-power2} are much smaller and $F_9$ is almost a factor of 2 larger than any of the four frequencies predicted for the tubular FS sheet.\cite{Hebbache-FS}  We suggest that $F_9$ is most likely a second harmonic of $F_7$.  The remaining  4 frequncies $F_3 = 2983$~T, $F_4 = 3812$~T, $F_5 = 3957$~T, and $F_7 = 5697$~T can be compared to the theoretically predicted frequencies 2177~T, 3265~T, 3888~T, and 5291~T\@.\cite{Hebbache-FS}  The experimentally observed frequencies are similar to those predicted considering the unknown angular dependence of the frequencies arising from the tubular FS sheet.  Thus we can tentatively assign these observed frequencies to the quasi-two-dimensional tubular FS sheet.

The most intense frequencies in Fig.~\ref{Fig-power2}, $F_4$, $F_5$, $F_6$, and $F_8$, were used to estimate the effective masses by fitting the $T$ dependences of these frequencies by Eq.~(\ref{Eq-LK(T)}).  We obtain $m^*_4 = 0.87 m_{\rm e}$, $m^*_5 = 1.13 m_{\rm e}$, $m^*_6 = 0.92 m_{\rm e}$, and $m^*_8 = 0.96 m_{\rm e}$.  The corresponding theoretically predicted band masses are $m_{\rm b4} = 0.95 m_{\rm e}$, $m_{\rm b5} = 0.92 m_{\rm e}$, $m_{\rm b6} = 0.62 m_{\rm e}$, and $m_{\rm b8} = 0.45 m_{\rm e}$.  Using the expression $m^* = (1+\lambda_{\rm ep}) m_{\rm b}$, we estimate electron-phonon interaction constants $\lambda_4 \approx 0$, $\lambda_5 \approx 0.23$, $\lambda_6 \approx 0.48$, and $\lambda_8 \approx 1.13$.  It should be noted that the frequencies $F_6$ and $F_8$ arise from the deformed ellipsoidal FS sheets.  Thus, we find that $\lambda_{\rm ep}$ is larger for the ellipsoidal FS sheets compared to the quasi-two-dimensional tubular FS sheet.  This suggests that the superconductivity in OsB$_2$ is driven by the two deformed ellipsoidal FS sheets.  The average value of $\lambda_{\rm ep}$ estimated above from McMillan's formula Eq.~(\ref{EqMcMillan2}) was 0.4--0.5 which agrees with our inference that $\lambda_{\rm ep}$ is small on some FS sheets and is larger on others.  

The above experimental and theoretical SdH data are summarized in Table~\ref{SdHFreqs}.

\section{Summary and Conclusions}
\label{sec:CON}

\begin{table}
\caption{\label{PropSumm} Parameters describing the physical properties of OsB$_2$.  The symbols are superconducting transition temperature $T_{\rm c}$, electrical resistivity $\rho_b$ along the $b$~axis, estimated mean-free path along the $b$~axis just above $T_{\rm c}$, $\ell_b(2.25$~K), normal state magnetic susceptibility $\chi_\alpha$ ($\alpha = a,\ b,\ c$) and powder average $\bar{\chi}$ measured in $H = 3$~T, linear specific heat coefficient $\gamma$, Debye temperature $\Theta_{\rm D}$, electron-phonon coupling constant $\lambda_{\rm ep}$, bare band structure density of states derived from heat capacity measurements $N(\epsilon_{\rm F})$ for both spin directions, Fermi velocity $v_{\rm F}$, nearly isotropic upper critical magnetic field $H_{\rm c2}$, superconducting coherence length $\xi$, magnetic penetration depth $\lambda$, thermodynamic critical field $H_{\rm c}$, and Ginzburg-Landau parameter $\kappa$.}

\begin{ruledtabular}
\begin{tabular}{lc}
Quantity & value\\ \hline
$T_{\rm c}$ & 2.10(5)~K\\
$\rho_b(2.25$~K) & 1.55~$\mu\Omega$~cm\\
$\ell_b(2.25$~K) & $0.137~\mu$m\\
$\chi_a$(300 K) & ${\rm -3.9 \times 10^{-5}\ cm^3/mol}$ \\
$\chi_b$(300 K) & ${\rm -6.3 \times 10^{-5}\ cm^3/mol}$ \\
$\chi_c$(300 K) & ${\rm -3.2 \times 10^{-5}\ cm^3/mol}$ \\
$\bar{\chi}$(300 K) & ${\rm -4.50 \times 10^{-5}\ cm^3/mol}$ \\
$\gamma$ & 1.95(1)~mJ/mol~K$^2$\\
$\Theta_{\rm D}$ & 539(2)~K \\
$\lambda_{\rm ep}$ & 0.50 \\
$N(\epsilon_{\rm F})$ & 0.55 states/(eV f.u.)\\
$v_{\rm F}$ & $9.1\times10^7$~cm/s\\
$H_{\rm c2}(T = 0)$ & 186(4)~Oe\\
$\xi(T = 0)$ & 0.133(2)~$\mu$m\\
$\lambda(T = 0)$ & 0.300~$\mu$m \\
$H_{\rm c}(T = 0)$ & 153~Oe\\
$\kappa(T_{\rm c})$ & 2(1) \\
\end{tabular}
\end{ruledtabular}
\label{tabStruct}
\end{table}

We have grown high quality single crystals of OsB$_2$ using a novel Cu-B eutectic flux.  Measurements on these crystals confirm bulk superconductivity.  The crystallographic parameters of a single crystal are described above in Tables~\ref{CrystalData} and~\ref{atomicpositions} and Fermi surface properties in Table~\ref{SdHFreqs}.  The various parameters describing other normal and superconducting state properties are summarized here in Table~\ref{PropSumm}.  

The heat capacity measurements show some unusual behaviors.  The zero field anomaly at the superconducting transition $\Delta C/\gamma T_{\rm c} \approx 1.3$ is smaller than the weak-coupling BCS value of 1.43.  We suggest that this arises due to the two-gap nature of the superconductivity in OsB$_2$.  The occurrence of two superconducting gaps is supported by the anomalous temperature dependence of the penetration depth which could be fitted by the new $\gamma$ model for multi-gap superconductors\cite{Kogan2009} with the magnitudes of the two gaps being $\Delta_1(T=0)/k_{\rm B}T_{\rm c} = 1.90$ and $\Delta_2(T=0)/k_{\rm B}T_{\rm c} = 1.25$, respectively.  The zero-temperature upper critical field was determined to be $H_{\rm c2}(0) = 186$~Oe.  Four estimates of the Ginzburg-Landau parameter gave $\kappa(T_{\rm c}) \sim 1$--3 and thus indicate that OsB$_2$ is a small-$\kappa$ Type-II superconductor.  We observed an anomalous increase in the heat capacity jump at $T_{\rm c}$ measured in a finite magnetic field $H$.  For example, at $H \approx 25$~Oe, $\Delta C/\gamma T_{\rm c} \approx 2.5$.  This anomalous increase in $\Delta C/\gamma T_{\rm c}$ was confirmed for two batches of crystals.

The high quality of the crystals made it possible for us to study the anisotropy of the Fermi surface (FS) of OsB$_2$ by measuring Shubnikov-de Haas quantum oscillations via contactless rf skin depth measurements.  Some experimentally observed frequencies could be assigned to those predicted theoretically.  The effective masses estimated for the two deformed ellipsoidal FS sheets are larger than the predicted band masses and suggest a large electron-phonon coupling constant $\lambda_{\rm ep} \sim 0.5$--1 for these FS sheets.  A much smaller value of $\lambda_{\rm ep}$ was found for the third quasi-two-dimensional tubular FS sheet.  These results suggest that the superconductivity  in OsB$_2$ is driven by the two ellipsoidal FS sheets.  This would also explain the negligible anisotropy in the measured upper critical fields between the three crystallographic directions.

\begin{acknowledgments}
Work at the Ames Laboratory was supported by the Department of Energy-Basic Energy Sciences under Contract No. DE-AC02-07CH11358.  R.P.\ also acknowledges support from NSF Grant number DMR-05-53285 and from the Alfred P. Sloan Foundation.
\end{acknowledgments}


\begin{thebibliography}{99}
\bibitem{SHELXTL} All software and sources of the scattering factors are contained in the SHELXTL (version 5.1) program library (G. Sheldrick, Bruker Analytical X-Ray Systems, Madison, WI).
\end{thebibliography}
\end{document}